%% file: arxiv-v1.tex
\documentclass[aps,prl,twocolumn,superscriptaddress,groupedaddress,preprintnumbers]{revtex4}

\usepackage{graphicx}  
\usepackage{dcolumn}   
\usepackage{bm}        
\usepackage{amssymb}   
\usepackage{xcolor}

\usepackage[hidelinks]{hyperref}
\usepackage{amsmath}

\usepackage{subcaption}
\usepackage{comment}

\newcommand{\pythia}{\textsc{Pythia}}

\preprint{MIT-CTP/6018}

\begin{document}


\title{A Breath of Fresh Air for Moli\`ere:\\
Detecting Moli\`ere Scattering using Jet Substructure Observables in Oxygen Collisions}
\author{Arjun Srinivasan Kudinoor}
\email{kudinoor@mit.edu}
\affiliation{%
 Center for Theoretical Physics --- a Leinweber Institute, Massachusetts Institute of Technology, Cambridge, MA 02139
}%

\author{Arthur Yi-Ting Lin}
\email{arthur72@mit.edu}
\affiliation{%
 Massachusetts Institute of Technology, Cambridge, MA, 02139 }%

\author{Daniel Pablos}
\email{pablosdaniel@uniovi.es}
\affiliation{Departamento de F\'isica, Universidad de Oviedo, Avda. Federico Garc\'ia Lorca 18, 33007 Oviedo, Spain}
\affiliation{Instituto Universitario de Ciencias y Tecnolog\'ias Espaciales de Asturias (ICTEA), Calle de la Independencia 13, 33004 Oviedo, Spain}

\author{Krishna Rajagopal}
\email{krishna@mit.edu}
\affiliation{%
 Center for Theoretical Physics --- a Leinweber Institute, Massachusetts Institute of Technology, Cambridge, MA 02139
}%


\begin{abstract}
Ultra-relativistic oxygen-oxygen (OO) collisions are a promising arena in which to probe rare, large-angle, high momentum-transfer $2\rightarrow2$ Moli\`ere scatterings between energetic jet partons and quasiparticles in quark-gluon plasma (QGP).
As a jet propagates through the droplet of QGP formed in the same collision, its constituents lose energy to and excite wakes in the medium, and may scatter off quark- and gluon-like quasiparticles in QGP. Using the hybrid strong/weak coupling model, we show that including Moli\`ere scatterings between jet partons and medium quasiparticles is essential to reproduce recent CMS measurements of charged-particle suppression in OO collisions with this model. We then present the first theoretical study of how jet-medium interactions modify the internal structure of jets in OO collisions. We find that Moli\`ere scatterings broaden the Soft Drop splitting angle $R_g$, enhancing the population of $R=0.4$ and $R=0.8$ jets with $R_g\gtrsim0.2$ in OO collisions relative to pp collisions. Energy-energy correlators (EECs) provide a complementary probe, exhibiting enhanced large-angle correlations within jets due to jet-induced wakes and Moli\`ere scattering. In both cases, we propose an experimental measurement where the relevant OO/pp ratio can, if enhanced above unity in future data as in our calculations, be a distinctive, model-independent, detection
of hard scattering off QGP quasiparticles.
We furthermore use our calculations of EECs to
show how the angular scale corresponding to the deflection of jet or medium partons by Moli\`ere scattering is imprinted
in the EEC for jets with radius $R_{\rm jet}\sim0.8$ in OO collisions.
These results demonstrate that jet substructure measurements in OO collisions are promising avenues to probe the quasiparticles that emerge at short distances within an otherwise strongly coupled medium.
\end{abstract}

\maketitle


\textit{Introduction.} --- Viewed at length-scales of order the inverse of its temperature, QGP behaves as a strongly coupled liquid. When it is probed with high enough momentum transfer, though, asymptotic freedom mandates the presence of quark- and gluon-like quasiparticles.
Energetic partons within jets in light- and heavy-ion collisions can trigger high momentum exchanges with quasiparticles in QGP, making jets a compelling probe with which to study the microscopic structure of this strongly coupled liquid.
In this Letter, we investigate  rare, perturbative, high momentum-transfer, $2 \rightarrow 2$ Moli\`ere scattering between jet partons and QGP quasiparticles 
in ultra-relativistic oxygen--oxygen (OO) collisions. Recent OO collisions at the LHC and RHIC, with  first measurements of hard probes 
already appearing~\cite{Strangmann2025,ATLAS:2025ooe,CMS:2025bta}, are a promising 
arena for detecting and studying such scatterings. Since  the QGP droplets formed in OO collisions~\cite{Huang:2023qm,ALICE:2025luc,CMS:2025tga,ATLAS:2025nnt} 
are typically smaller than in PbPb collisions, jet observables benefit from a reduced contribution of strongly coupled energy loss, 
allowing effects of weakly coupled elastic collisions like Moli\`ere scattering 
to stand out.  

We first introduce the hybrid strong/weak coupling model of jet quenching (aka the Hybrid Model). 
We then show that Hybrid Model calculations agree with CMS measurements~\cite{CMS:2025bta} of charged-particle suppression in OO collisions only if Moli\`ere scatterings are included in the model. This motivates studying how Moli\`ere scatterings modify the internal structure of jets, specifically two jet substructure observables: the Soft Drop angle $R_g$ and energy-energy correlator (EEC). We find that large-angle deflections of 
jet partons
due to Moli\`ere scattering results in a broadening of $R_g$. Furthermore, we show that EECs exhibit enhanced large-angle correlations driven by Moli\`ere scatterings that can be separated from effects of the wakes that jets excite in the medium by imposing a track cut. 
This
first theoretical analysis of jet substructure observables in OO collisions 
demonstrates the promise of two observables for detecting and studying Moli\`ere scattering off quasiparticles in QGP. We provide a roadmap via which experimental measurements may yield distinctive model-independent signatures of this phenomenon and can quantify the typical angular scale of the resulting deflection of jet or QGP partons. Success in this regard would realize a vision long identified as a central goal of the field~\cite{Aprahamian:2015qub}.

\textit{The Hybrid Model.} --- The Hybrid Model 
describes the production, evolution, and modification of parton showers as they propagate through the droplets of QGP formed in light- and heavy-ion collisions. The production and subsequent evolution of these parton showers are determined by high-virtuality, perturbative, QCD evolution, as implemented in {\pythia}~8~\cite{Sjostrand:2007gs,Sjostrand:2014zea}. 
In the Hybrid Model, each parton in a jet shower loses energy to the plasma via a holographically derived formula for $dE/dx$~\cite{Chesler:2014jva,Chesler:2015nqz} implemented in Refs.~\cite{Casalderrey-Solana:2014bpa,Casalderrey-Solana:2015vaa,Casalderrey-Solana:2016jvj,Hulcher:2017cpt,Casalderrey-Solana:2018wrw,Casalderrey-Solana:2019ubu,Hulcher:2022kmn,Bossi:2024qho,Kudinoor:2025ilx,Kudinoor:2025gao,Beraudo:2025nvq,Hulcher:2026dht}. The strength of the interaction between a jet parton and the QGP is governed by a dimensionless parameter $\kappa_{\rm sc}$; for illustration, a parton with initial energy $E_{\rm in}$ thermalizes over a distance $x_{\rm stop}=E_{\rm in}^{1/3}/(2\kappa_{\rm sc}T^{4/3})$ if it does not split first. 
We treat $\kappa_{\rm sc}$ as a phenomenological parameter in the Hybrid Model, fitting it to jet and high-$p_T$ hadron suppression data in PbPb collisions as first done in Refs.~\cite{Casalderrey-Solana:2018wrw,Hulcher:2026dht}. As described in the Supplemental Material~\cite{supplement}, when Moli\`ere scatterings are (not) included, we choose $\kappa_{\rm sc}=0.335 \, (0.37)$. These choices
were made using Hybrid Model calculations
for PbPb collisions only.
These values of $\kappa_{\rm sc}$ differ from those in Refs.~\cite{Casalderrey-Solana:2018wrw,Hulcher:2026dht} because 
previous studies
used event-averaged hydrodynamic  profiles, whereas in this study we use state-of-the art, event-by-event hydrodynamic  profiles~\cite{Mantysaari:2025tcg} because the
event-by-event fluctuations in the shapes of  droplets of QGP in collisions of smaller nuclei are significant.
(See Supplemental Material~\cite{supplement} for details.)


The energy and momentum lost by a jet parton is deposited into the plasma, exciting a hydrodynamic wake in the expanding, flowing, cooling droplet of liquid QGP.
In the Hybrid Model, jet wakes are implemented by generating soft hadrons according to a spectrum determined by applying the Cooper--Frye prescription to the jet-induced perturbation of the stress-energy tensor of the liquid QGP~\cite{Casalderrey-Solana:2016jvj}.
The wake-spectrum and the assumptions in its calculation are discussed in Refs.~\cite{Casalderrey-Solana:2016jvj,Casalderrey-Solana:2020rsj,Bossi:2024qho,Kudinoor:2025ilx}.

Elastic $2 \rightarrow 2$ Moli\`ere scatterings with high momentum transfer are a different, intrinsically weakly coupled, channel for energetic jet partons to interact with the medium. In the Hybrid Model~\cite{Hulcher:2026dht}, a jet parton that scatters is deflected, kicking a medium parton, which recoils. As both of these partons propagate further through the medium they lose energy and momentum, excite wakes in the medium, and can re-scatter~\cite{Hulcher:2026dht, Hulcher:2022kmn}.
As in Ref.~\cite{Hulcher:2026dht}, for simplicity at present we compute the probability that a jet parton scatters off a massless quark or gluon quasiparticle drawn from a thermal distribution and we enforce large momentum exchange during the perturbative scattering process by requiring that the Mandelstam variables $t$ and $u$ satisfy $|t|, \, |u| > a m_D^2$, choosing the threshold parameter $a=10$. Here, 
the squared Debye mass is $m_D^2 = g_s^2T^2\bigl(N_c+N_f/2\bigr)/3$ with $N_c=N_f=3$. As in Ref.~\cite{Hulcher:2026dht}, both here and in the 
matrix elements for elastic scattering we shall take $g_s=2.25$, corresponding to $\alpha_s\simeq 0.4$. 
As in Refs.~\cite{Casalderrey-Solana:2016jvj,Hulcher:2026dht}, we model
transverse momentum kicks below the threshold $|t|, \,|u|<a m_D^2$ as soft Gaussian transverse momentum broadening with a jet parton that travels $\delta x$ after a splitting picking up transverse momentum $\langle k_\perp^2\rangle=K T^3 \delta x$,
where we choose the parameter $K=15$, as described in Ref.~\cite{Hulcher:2026dht}.



\begin{figure}[t]
\includegraphics[width=\linewidth]{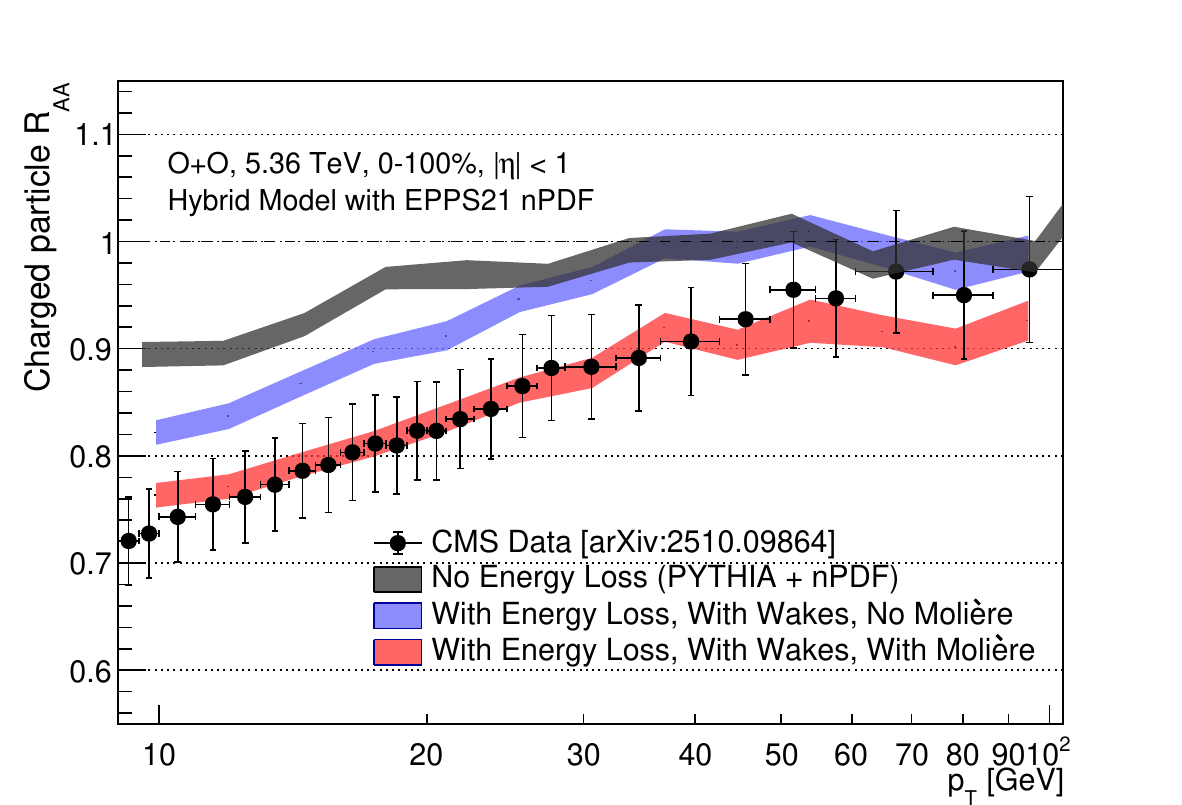}
\caption{
    \label{fig:ch-had-raa}Hybrid Model calculations of $R_{\rm AA}$ of charged hadrons with $|\eta| < 1$ in OO collisions versus $p_T$ 
    without energy loss or Moli\`ere scattering (gray), with energy loss and wakes (blue), and with energy loss, wakes, and Moli\`ere scatterings (red). Points denote CMS 
    measurements~\cite{CMS:2025bta}, with statistical, systematic, and normalization errors added in quadrature.
}
\end{figure}

\textit{High-$p_T$ hadron suppression in OO Collisions.} --- We begin in Fig.~\ref{fig:ch-had-raa} by comparing Hybrid Model calculations of the suppression $R_{\rm AA}$ 
of charged-hadron production in OO collisions
vs.~hadron transverse momentum $p_T$ to recent CMS measurements~\cite{CMS:2025bta}. 
($R_{\rm AA}$ is the number of charged hadrons in a $p_T$-bin in OO collisions relative to that in the corresponding number of pp collisions.) Unless stated otherwise, calculations in this Letter were performed for minimum-bias (0-100\% centrality) OO collisions with collision energy $\sqrt{s_{\rm NN}} = 5.36$~TeV. 

Fig.~\ref{fig:ch-had-raa} shows  $R_{\rm AA}$ for charged hadrons with $|\eta| < 1$ versus $p_T$. The colored bands depict Hybrid Model calculations, while the point markers denote CMS data~\cite{CMS:2025bta}. The gray band, which only includes  initial-state effects from the EPPS21 nuclear parton distribution functions (nPDFs)~\cite{Eskola:2021nhw} and excludes all effects of quenching, fails to describe the CMS data. Including the effects of strongly coupled energy loss and jet wakes (blue) yields modest suppression, but agreement with CMS data is achieved only when Moli\`ere scatterings are also included (red).  This 
constitutes evidence for the 
presence and importance of Moli\`ere scattering, but this evidence is far from model-independent: it depends on EPPS21 modeling of nPDFs 
(see Refs.~\cite{Brewer:2021tyv,Paakkinen:2021jjp,Gebhard:2024flv,Mazeliauskas:2025clt,Jonas:2026yoz,Koley:2026gdn} for investigations of nPDF effects), and our modeling of parton energy loss 
(see Refs.~\cite{Katz:2019qwv,Huss:2020dwe,Huss:2020whe,Zakharov:2021uza,Xie:2022fak,Ke:2022gkq,Ogrodnik:2024qug,vanderSchee:2025hoe,Pablos:2025cli,Faraday:2025pto,Faraday:2025prr} 
for other treatments of parton energy loss in OO collisions),
jet wakes, and Moli\`ere scattering itself.

\begin{figure*}
    \centering
    \begin{subfigure}[t]{0.45\textwidth}
        \centering
        \includegraphics[width=\textwidth]{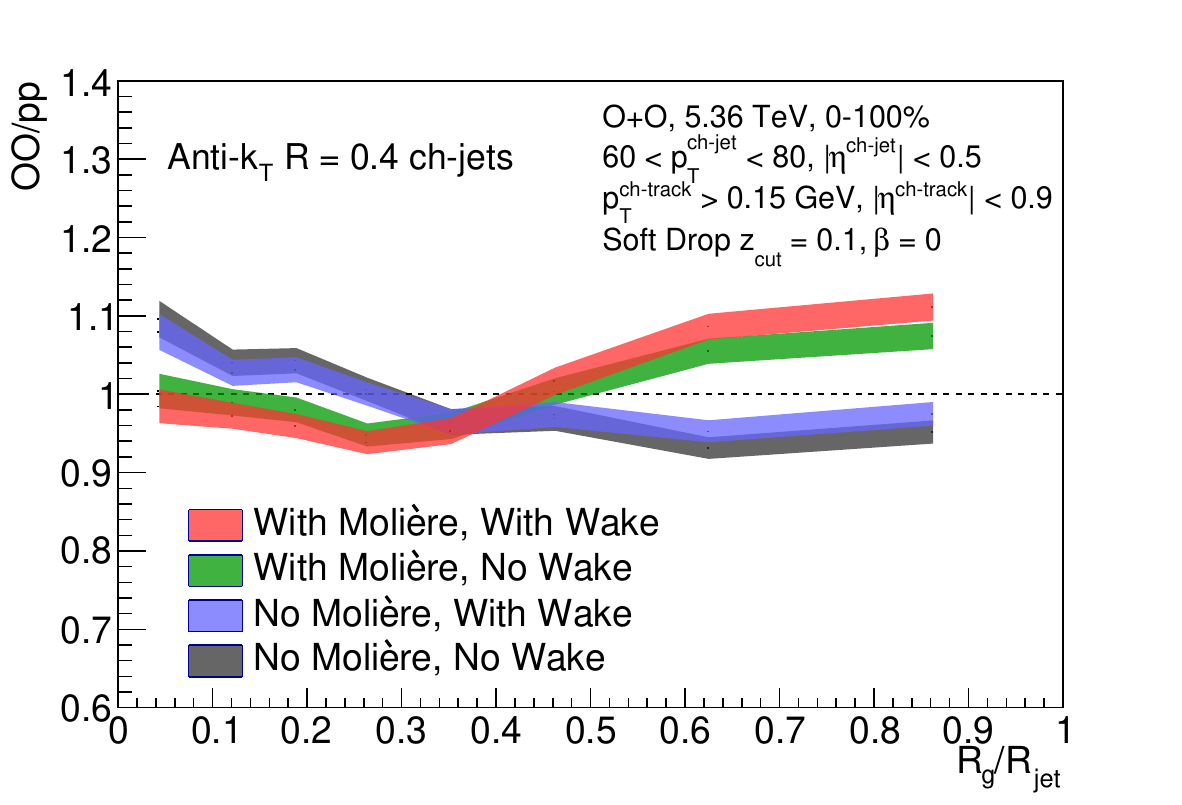}
    \end{subfigure}%
    ~ 
    \begin{subfigure}[t]{0.45\textwidth}
        \centering
        \includegraphics[width=\textwidth]{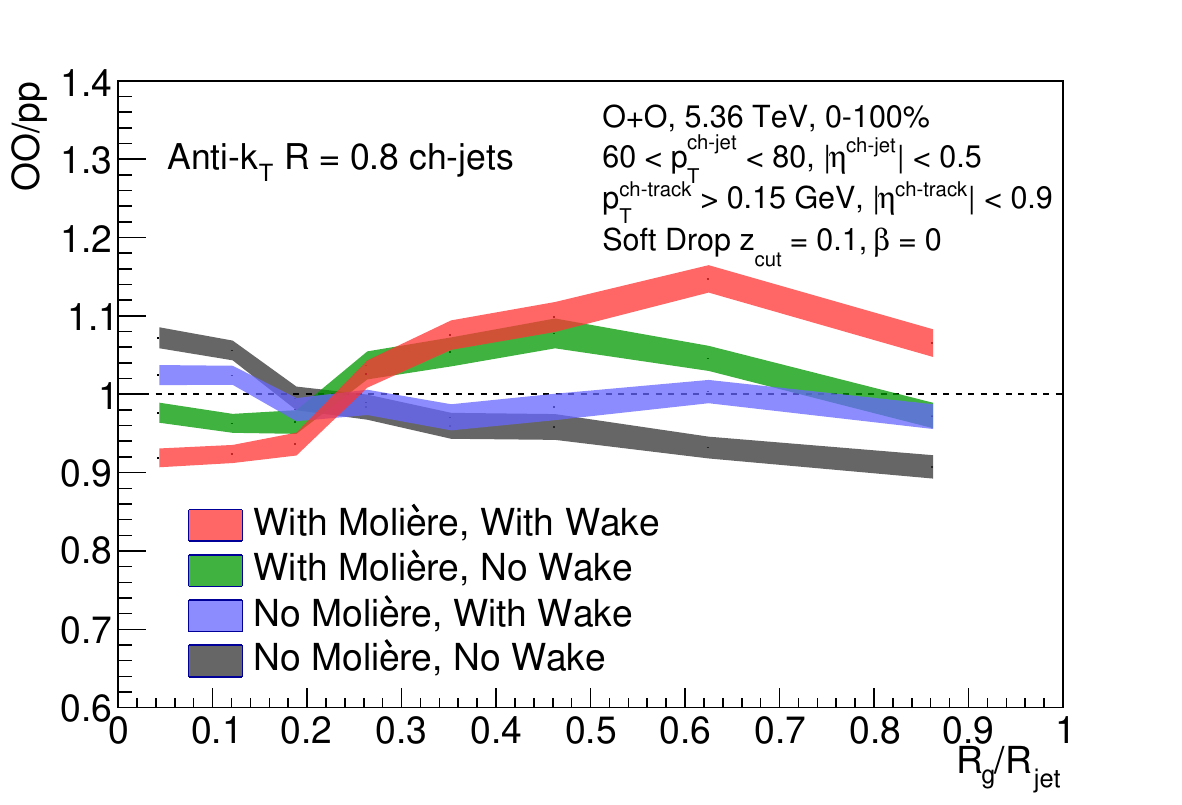}
    \end{subfigure}
    \caption{Hybrid Model calculations of the OO/pp ratio of the $R_g$/$R_{\rm jet}$ distributions in $R_{\rm jet} = 0.4$ (left) and $R_{\rm jet} = 0.8$ (right) jets with $40<p_T^{\rm jet}<60$~GeV, calculated using charged particles with $p_T^{\rm ch~track} > 150$~MeV,  Soft Drop grooming parameters $z_{\rm cut} = 0.1$ and $\beta = 0$, and OO and pp distributions  normalized by the number of jets selected.
    }
    \label{fig:rg}
\end{figure*}

Moli\`ere scatterings occur in both PbPb  and OO collisions, but their relative importance is amplified in OO collisions. This follows from 
a parametric argument:
collisional energy loss due to weakly coupled Moli\`ere scatterings kicking partons out of the jet cone scales linearly with the in-medium path length $L$ traversed by jet partons, 
whereas strongly coupled energy loss scales as $L^3$ for $L \ll x_{\rm stop}$ and more strongly with $L$ for  partons that thermalize~\cite{Chesler:2014jva,Chesler:2015nqz}. Since OO collisions produce smaller QGP droplets than PbPb collisions,
the contribution that Moli\`ere scatterings 
make to suppressing charged hadron production
is larger relative to 
that of energy loss in OO collisions.
This makes the lightest ion collisions in which jet quenching is apparent --- which today means OO collisions --- the best arena in which to
detect, isolate, quantify, and analyze the effects of Moli\`ere scatterings.

Observables
like hadron $R_{\rm AA}$, jet $R_{\rm AA}$ and dijet asymmetry (see the Supplemental Material~\cite{supplement}) 
indicate
the degree to which energetic partons and jets are quenched in OO collisions. To understand how 
energy and momentum are redistributed within 
a jet and how jet partons resolve the microscopic structure of QGP,
we study the impacts of Moli\`ere scatterings and jet wakes on two jet substructure observables: the Soft Drop splitting angle $R_g$ and energy-energy correlators (EECs).

\textit{Soft Drop $R_g$ Modifications.} --- The \textit{Soft Drop} grooming algorithm~\cite{Larkoski:2014wba} is designed to identify the first hard splitting in a jet; for jets in heavy- or light-ion collisions, it also grooms away most soft hadrons originating from their wakes~\cite{Casalderrey-Solana:2019ubu}. Since Moli\`ere scattering imparts rare, but sizable, 
momentum kicks to jet partons~\cite{Kurkela:2014tla} and the partons from the medium that they strike, 
it broadens the angular separation of resolved splittings within jets. 
This makes Soft Drop a powerful tool for detecting Moli\`ere scattering of partons in a jet traversing QGP.
The Soft Drop algorithm is as follows:  first, reconstruct jets using the anti-$k_t$ algorithm~\cite{Cacciari:2008gp} with radius parameter $R_{\rm jet}$; next, recluster the jet constituents with the Cambridge--Aachen algorithm~\cite{Dokshitzer:1997in,Wobisch:1998wt}, whose clustering history follows the angular structure of the parton shower; then, the Soft Drop algorithm steps through the history and selects the first splitting to satisfy the condition
$z>z_{\rm cut}(R_{12}/R_{\rm jet})^\beta$,
where we choose the parameters
$z_{\rm cut} = 0.1$ and $\beta = 0$,
$z$ is the fraction of transverse momentum carried by the subleading prong in the splitting, and $R_{12}$ is the angular separation between the two prongs. The Soft Drop angle $R_g$ is defined
as the $R_{12}$ of the first splitting that fulfills this condition.



Although Moli\`ere scattering within a jet tends to broaden its $R_g$, this need not push the $R_g$ distribution of jets with a given $p_T$ to larger $R_g$ in PbPb collisions than in pp collisions.
Because the jet production rate falls rapidly with increasing jet $p_T$, selecting jets with a given $p_T$ in PbPb collisions 
yields a sample that is biased against jets that start out with much higher $p_T$ and lose a lot of energy, and biased towards jets that survive the QGP having lost little energy.
This 
favors  jets with fewer and narrower splittings~\cite{Milhano:2015mng,Casalderrey-Solana:2016jvj,Casalderrey-Solana:2018wrw}, with a smaller $R_g$, and depresses the PbPb/pp ratio of $R_g$ distributions 
at large $R_g$~\cite{Casalderrey-Solana:2019ubu,Kudinoor:2025gao,Hulcher:2026dht} as seen in experimental data~\cite{ALargeIonColliderExperiment:2021mqf, CMS:2024zjn}.
This confounding effect reduces, and  in PbPb collisions may overwhelm, the effect of Moli\`ere scatterings on the $R_g$ distribution~\cite{CMS:2024zjn,Hulcher:2026dht}. Because OO collisions produce smaller droplets of QGP than PbPb collisions, 
high-$p_T$ particles lose less energy in OO collisions. 
This should reduce the confounding effects of selection bias due to energy loss, making OO collisions a compelling arena in which to study the effects of Moli\`ere scattering.

\begin{figure*}
    \centering
    \begin{subfigure}[t]{0.45\textwidth}
        \centering
        \includegraphics[width=\textwidth]{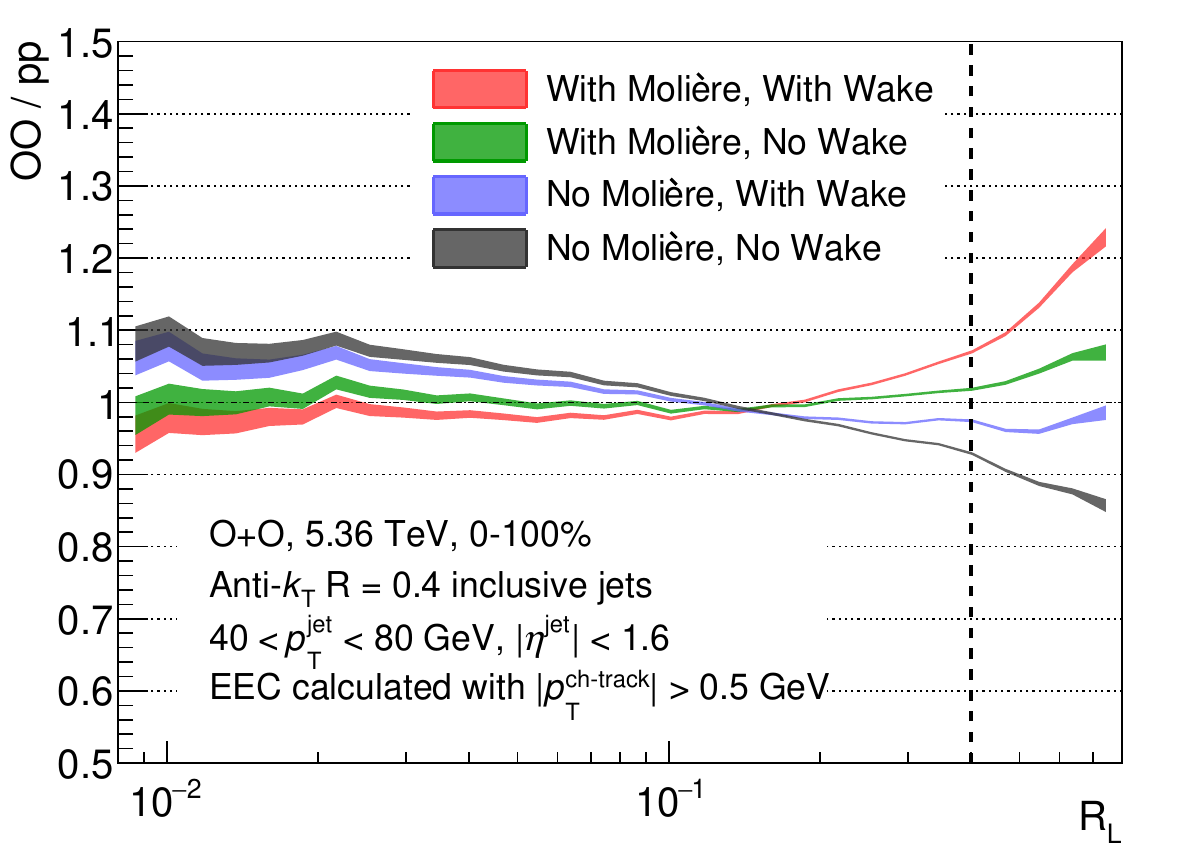}
    \end{subfigure}%
    ~ 
    \begin{subfigure}[t]{0.45\textwidth}
        \centering
        \includegraphics[width=\textwidth]{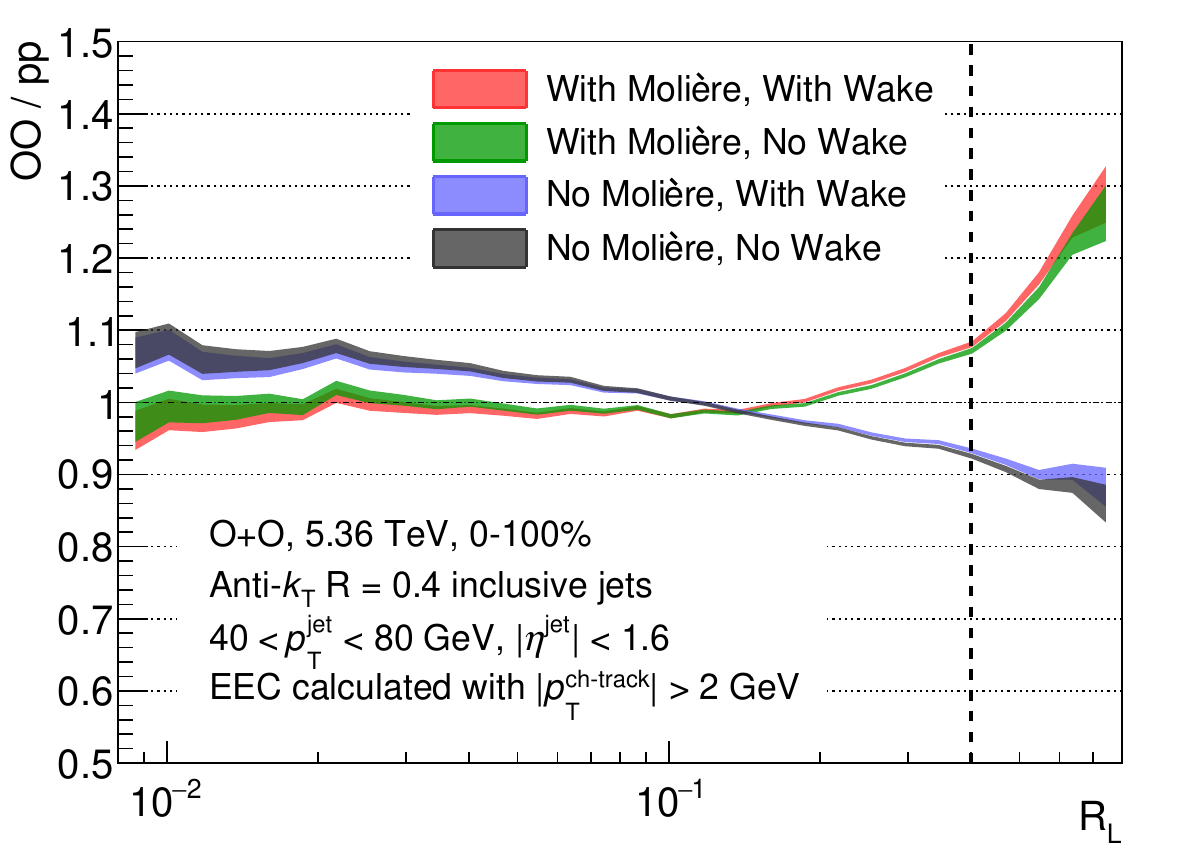}
    \end{subfigure}
    ~
    \begin{subfigure}[t]{0.45\textwidth}
        \centering
        \includegraphics[width=\textwidth]{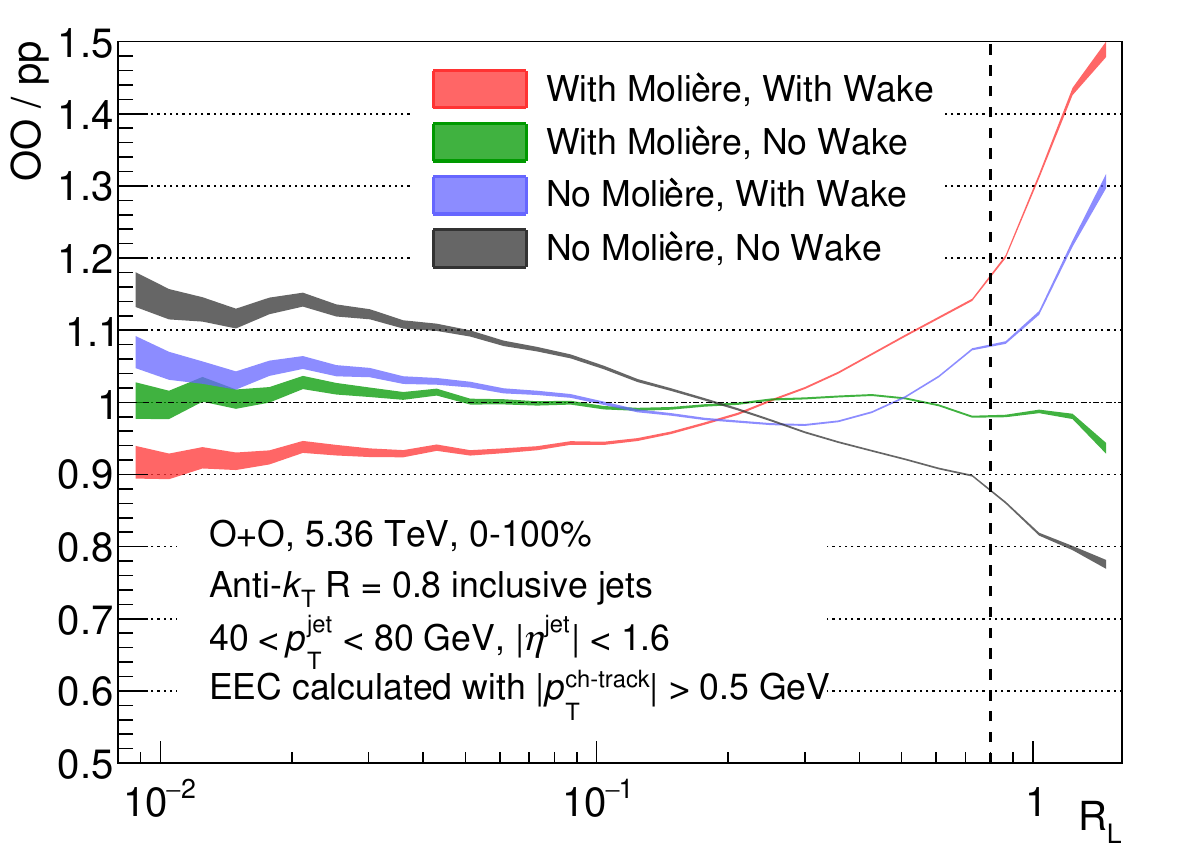}
    \end{subfigure}%
    ~ 
    \begin{subfigure}[t]{0.45\textwidth}
        \centering
        \includegraphics[width=\textwidth]{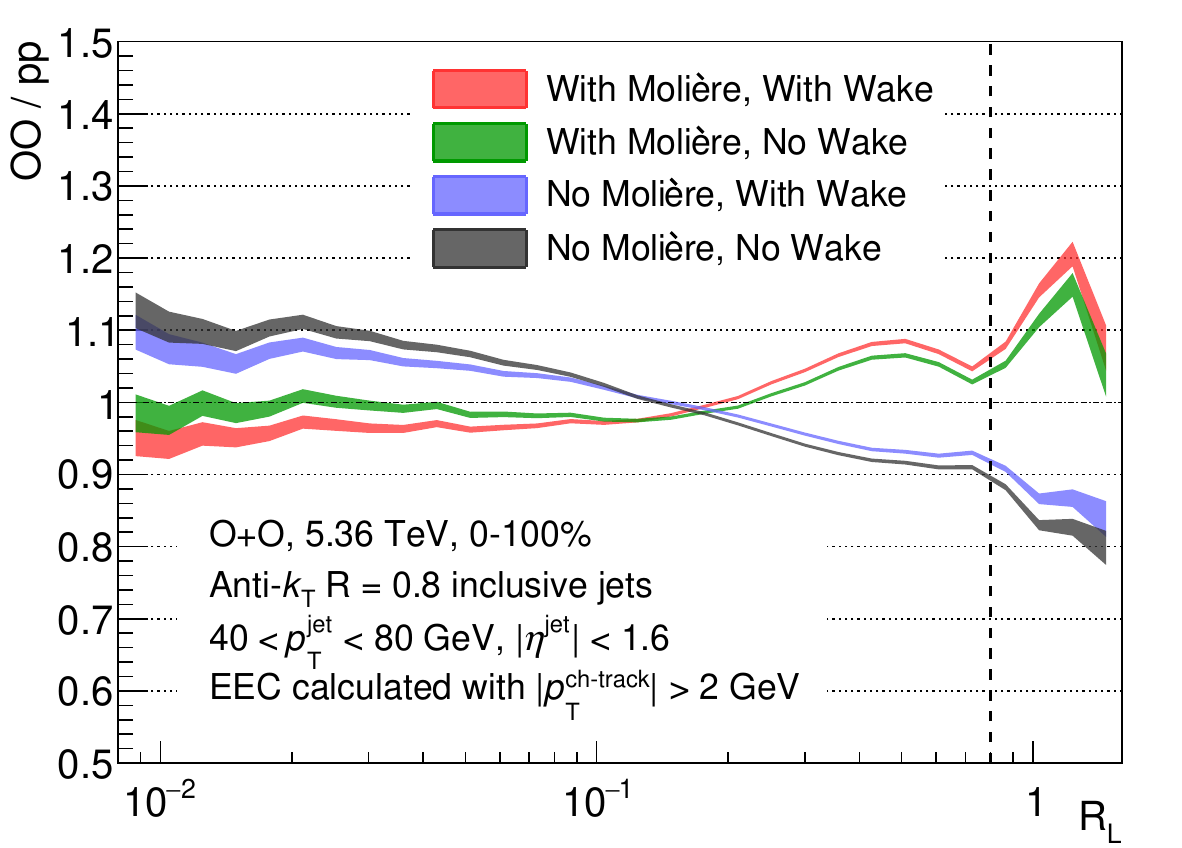}
    \end{subfigure}
    \caption{
    Hybrid Model calculations of OO/pp ratios of EECs for $R_{\rm jet} = 0.4$ (top panels) and $R_{\rm jet} = 0.8$ (bottom panels) jets with $40 < p_T < 80$ GeV versus $R_L$. EECs are calculated using charged hadrons with $p_T>0.5\, {\rm GeV}$ (left panels), and $p_T>2\, {\rm GeV}$ (right panels). To the right of the dashed vertical lines at $R_L = R_{\rm jet}$, EECs probe correlations between points separated by an $R_L$ that is greater than the jet radius and less than its diameter.
    } 
    \label{fig:eec min bias 40-80}
\end{figure*}

Fig.~\ref{fig:rg} shows Hybrid Model calculations of the ratio of the number of jets with a specified scaled Soft Drop angle $R_g/R_{\rm jet}$ in OO collisions to that in pp collisions, for $R_{\rm jet} = 0.4$ (left panel) and $R_{\rm jet}=0.8$ (right panel) jets with $60 < p_T^{\rm ch\text{-}jet} < 80$ GeV reconstructed from charged particles with $p_T > 150$ MeV. 
The four colored bands represent Hybrid Model calculations with wakes and Moli\`ere scatterings included/excluded.
The black and blue  bands (no Moli\`ere scattering) show modest effects of selection bias due to energy loss; comparison to analogous calculations for PbPb collisions~\cite{Hulcher:2026dht} confirms that the effects of 
energy loss are much less in OO collisions here. 
(The same conclusion can be drawn by comparing $R_{\rm AA}$ for charged hadrons in OO and PbPb collisions.) 
Comparing the blue and black bands in Fig.~\ref{fig:rg} confirms that the Soft Drop procedure grooms away most soft hadrons  from the freezeout of jet wakes, 
essentially eliminating their effects 
in jets with $R_{\rm jet}=0.4$ and, for $R_g\lesssim 0.4$, in jets with $R_{\rm jet}=0.8$.  The effects of jet wakes
seen for $R_g\gtrsim 0.4$ in the larger radius 
jets arise because
more of the hadrons from the wake remain in the jet cone. 


We find in Fig.~\ref{fig:rg} that including Moli\`ere scatterings leads to a clear enhancement
in the population of jets with $R_g\gtrsim 0.2$
in OO collisions relative to pp collisions, for jets with  $R_{\rm jet}=0.4$ and $R_{\rm jet}=0.8$.
The results of our calculations of $R_g$ are particularly striking for $R_{\rm jet}=0.4$ jets, where jet wakes have little effect.  
In this case, seeing an OO/pp ratio above unity at sufficiently large $R_g$ is a distinctive consequence
of Moli\`ere scattering.  In inclusive jets in PbPb collisions, this effect is overwhelmed by the countervailing effect of selection bias due
to energy loss~\cite{Hulcher:2026dht}; not here!
Measuring this in experimental data would be a distinctive, model-independent, signature of hard scattering 
of jet partons off QGP quasiparticles.



\textit{EEC Modifications.} --- 
Energy-energy correlators are complementary observables for studying how Moli\`ere scatterings modify both the hard shower and the soft wakes that partons in the shower excite in the medium~\cite{Andres:2022ovj,Andres:2023xwr,Andres:2023ymw,Yang:2023dwc,Barata:2023bhh,Barata:2023zqg,Andres:2024ksi,Andres:2024hdd,Xing:2024yrb,Fu:2024pic,Andres:2024xvk,Barata:2024ukm,Bossi:2024qho,Apolinario:2025vtx,Barata:2025fzd,Barata:2025zku,Ke:2025ibt,Liu:2025ufp,Andres:2025yls}. We study the two-point EEC, defined in Ref.~\cite{CMS:2025ydi} as
\begin{equation} \label{eq:eec}
    {\rm EEC}(R_L)
= \frac{1}{\mathcal{N}}\,\frac{1}{\delta r}
\sum_{\rm jets}
\, \, \sum_{{\rm pairs} \in [R_{L,a},\,R_{L,b}]} p_{T,i}\, p_{T,j},
\end{equation}
where $(i,j)$ refers to pairs of charged particles, $R_L \equiv \Delta r_{i,j} = \sqrt{(\eta_i - \eta_j)^2 + (\phi_i - \phi_j)^2}$ is the angular separation between particles $i$ and $j$, $R_{L, a}$ and $R_{L, b}$ are $R_L$ bin boundaries,  $\delta r \equiv R_{L, b} - R_{L, a}$ is the bin width,
and the normalization  is $\mathcal{N}=\int_{0.008}^{R_{\rm jet}}{\rm EEC}(R_L)d R_L$.
As in Ref.~\cite{CMS:2025ydi}, we first reconstruct jets in each collision event 
using the anti-$k_t$ algorithm, with E-scheme recombination~\cite{Cacciari:2011ma}, and then
identify the axis of each jet by applying the winner-take-all recombination algorithm~\cite{Bertolini:2013iqa, Larkoski:2014uqa} to its constituents. Finally, we identify all charged-particle tracks in the event (whether or not they are part of the reconstructed jet) that lie within an angular separation $R_{\rm jet}$
of the jet axis and use these tracks to compute  EEC$(R_L)$. Experimental measurements of EECs in PbPb~\cite{CMS:2025ydi, Rai:2025qm} and pPb collisions~\cite{Nambrath:2025ttz} have been reported.


Fig.~\ref{fig:eec min bias 40-80} shows Hybrid Model calculations of the OO/pp ratios of EEC$(R_L)$ for anti-$k_t$ $R_{\rm jet}=0.4$ jets (top) and 
$R_{\rm jet} = 0.8$ jets (bottom) with $40 < p_T < 80$ GeV. The EECs are calculated using charged-particle tracks with $p_T>0.5$ GeV (left) and $p_T>2$ GeV (right). 
The four colored bands represent Hybrid Model calculations with the wake and Molière scattering included/excluded. 
We see immediately in the right panels that imposing a track cut
of $p_T>2$~GeV almost completely eliminates the effects of jet wakes, as it serves to eliminate almost all of the soft hadrons from jet wakes from the analysis. 

In analyzing the effects of Moli\`ere scattering on EECs, it will be helpful to consider 
three regions:
$R_L \lesssim 0.2$, $0.2 \lesssim R_L < R_{\rm jet}$, and $R_L > R_{\rm jet}$.
The region $R_L \lesssim 0.2$ probes small-angle correlations  
dominated by hard particles in the cores of jets. 
In the absence of Moli\`ere scattering, the sample of jets with a given $p_T$ in OO collisions 
will be biased towards jets with narrower splittings than in pp collisions.
Correspondingly, the effect of selection bias due to energy loss seen in the black bands in all panels of Fig.~\ref{fig:eec min bias 40-80}
shows an enhancement of small-$R_L$
correlations in OO collisions relative to pp collisions.
%
%
Including either wakes (which broaden the soft shape of jets) or Moli\`ere scatterings (which broaden the angular distribution of semi-hard structures in jets) suppresses the EEC at small-$R_L$ and (see below) enhances it at larger $R_L$.


\begin{figure}
    \centering
\includegraphics[width = \linewidth]{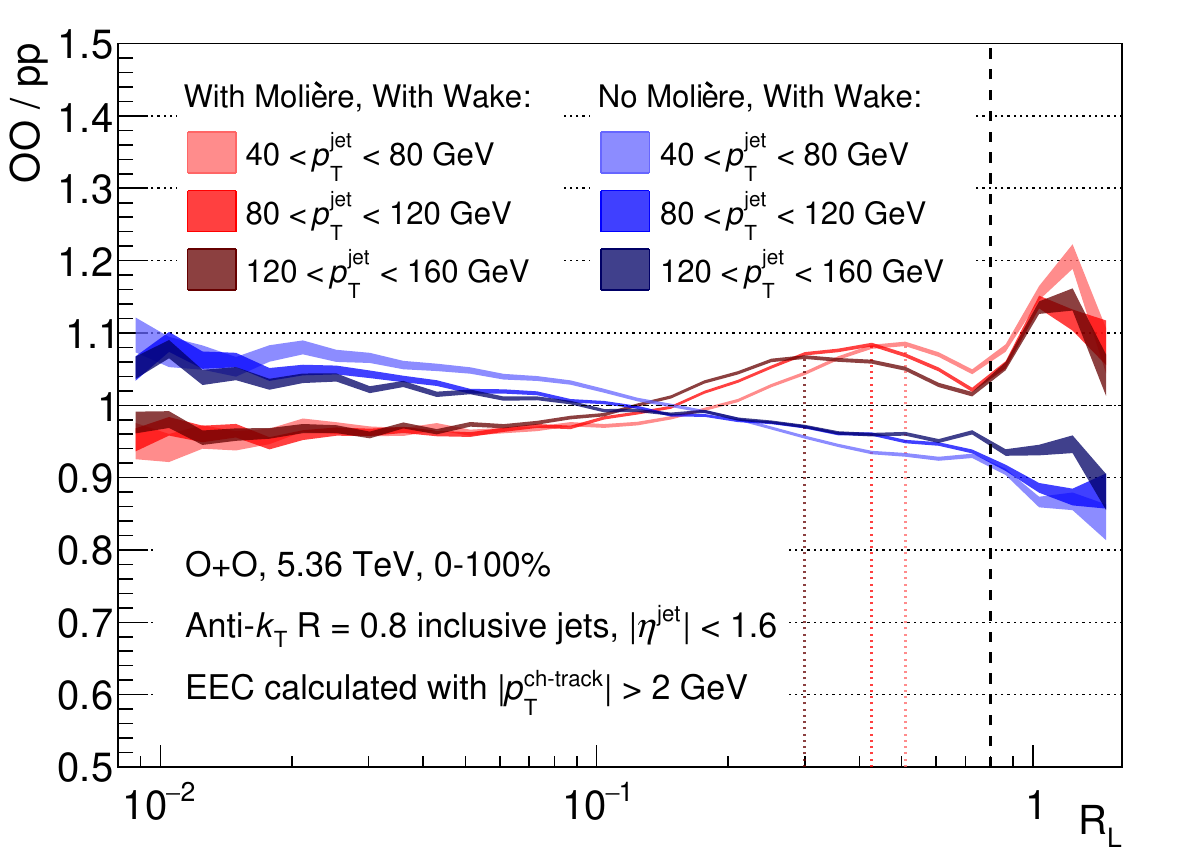}
    \caption{
    Hybrid Model calculations of OO/pp ratios of EECs with (red) and without (blue) Moli\`ere scattering for $R_{\rm jet} = 0.8$ jets with $p_T \in [40, 80]$ GeV, $[80, 120]$ GeV, and $[120, 160]$ GeV, calculated using charged hadrons with $p_T>2$ GeV. The dashed vertical line corresponds to $R_L = R_{\rm jet}$. Each dotted vertical line marks a local maximum of the EEC with $R_L < R_{\rm jet}$, whose location in $R_L$ encodes the typical angle of deflection due to Moli\`ere scatterings in the corresponding jet sample.
    }
    \label{fig:eec peaks}
\end{figure}

In the region $0.2 \lesssim R_L < R_{\rm jet}$, the EEC is dominated by correlations 
between one particle in the central core of a jet and another in its periphery~\cite{Bossi:2024qho}.
%
In the left panels of Fig.~\ref{fig:eec min bias 40-80}, we see significant enhancements to the EECs of jets in OO collisions in the region $0.2 \lesssim R_L < R_{\rm jet}$
arising from either jet wakes or Moli\`ere scattering, with the enhancement largest when both effects are included (which incorporates the soft particles populating the periphery of jets coming from the wakes excited by the outgoing particles after a Moli\`ere scattering).
%
We see in both right panels of Fig.~\ref{fig:eec min bias 40-80} that imposing $p_T^{\rm ch~track}> 2$ GeV eliminates essentially all effects of the soft hadrons from jet wakes.
With this track cut, an OO/pp ratio of EEC$(R_L)$
exceeding unity 
in the  $0.2 \lesssim R_L < R_{\rm jet}$ region 
would constitute a distinctive, model-independent,
experimental 
signature of Moli\`ere scattering, 
just as we found for the groomed $R_g$ observable in the left panel of Fig.~\ref{fig:rg} but here for jets with $R_{\rm jet}=0.8$ as well as $R_{\rm jet}=0.4$.



In the bottom-right panel of Fig.~\ref{fig:eec min bias 40-80}, there is an additional feature of Moli\`ere scattering, beyond just an enhanced large-angle correlation, imprinted on the EEC.
We see a 
local maximum --- a bump --- in the OO/pp ratio of EEC$(R_L)$ for $R_{\rm jet} = 0.8$ jets at an angle $R_L \simeq 0.5$.  We can see this bump only for jets with $R_{\rm jet}$ well above this angle.
Intuitively, this angle must correspond to the typical angle at which a parton in this sample of jets is deflected after a Moli\`ere scattering. 
To confirm this interpretation, in Fig.~\ref{fig:eec peaks} we look
at how the OO/pp ratios of EEC$(R_L)$ 
for $R_{\rm jet}=0.8$ jets
with the track cut $p_T^{\rm ch~track} > 2$ GeV
depend on jet $p_T$.
Since higher-momentum partons are more likely to scatter at smaller angles~\cite{Hulcher:2026dht}, the position of this bump in $R_L$ should decrease with increasing jet $p_T$, as is indeed apparent in the pink/red/brown curves in Fig.~\ref{fig:eec peaks}.
%
%
%
Observing this bump and its dependence on jet $p_T$ in experimental measurements of the OO/pp ratios of EEC$(R_L)$ would be additional 
exciting evidence for 
Moli\`ere scattering. And, such data would be key inputs to a future Bayesian uncertainty quantification that constrains Hybrid Model parameters like $\kappa_{\rm sc}$, $g_s$, $a$ and $K$.
%
%
These calculations 
were
performed in minimum-bias (0-100\% central) OO collisions; in the Supplemental Material~\cite{supplement}, we show  analogous calculations 
for different collision centralities.

Finally, the region $R_L > R_{\rm jet}$ probes  
correlations between points separated by such a large angle that neither point can be in the central core of the jet.
In the left panels of Fig.~\ref{fig:eec min bias 40-80}, where we include 
soft hadrons with $p_T^{\rm ch~track}>0.5$~GeV,
as previously predicted for PbPb collisions~\cite{Bossi:2025kac} we see very large contributions to the large-angle EEC arising from jet wakes, which populate the peripheral regions of the jet with soft hadrons. 
Looking at  $R_L>R_{\rm jet}$  in the right panels of Fig.~\ref{fig:eec min bias 40-80}, we see that Moli\`ere scattering
also populates the peripheral regions of the jet, in this case with hadrons with $p_T>2$~GeV.

\textit{Conclusions.} --- We have investigated how Moli\`ere scatterings between jet partons and QGP quasiparticles modify jet substructure in OO collisions. Because the QGP droplets produced in OO collisions are smaller than those in PbPb collisions, the effects of strongly coupled energy loss and jet-selection bias are reduced. As a result, OO collisions provide a particularly favorable environment in which to isolate and study Moli\`ere scatterings.

We showed that Moli\`ere scatterings are essential to describe CMS measurements~\cite{CMS:2025bta} of charged-particle suppression in OO collisions with the Hybrid Model and studied how they affect jet substructure via calculations of the Soft Drop angle $R_g$ and energy-energy correlators. We found that Moli\`ere scatterings broaden $R_g$, leading to an enhanced population of $R_{\rm jet} = 0.4$ and 
$0.8$ jets with $R_g \gtrsim 0.2$ in OO collisions relative to pp collisions.
EECs are complementary observables, exhibiting enhanced large-angle correlations within jets due to wakes and Moli\`ere scatterings. Restricting to  charged-particle tracks with $p_T > 2$ GeV isolates the effects of Moli\`ere scattering on EECs. With this kinematic restriction, 
the typical angle at which hard partons are deflected due to Moli\`ere scattering 
is manifest as an experimentally measurable
bump in the OO/pp EEC ratio in $R_{\rm jet}=0.8$ jets at an angle $R_L$ that decreases with increasing jet $p_T$.


Observing the Moli\`ere-scattering-induced broadening of the $R_g$ distribution, or enhancement of the EEC$(R_L)$ for particles with $p_T>2$~GeV at $R_L\gtrsim 0.2$, or both
would provide distinctive, model-independent, evidence of hard scatterings between jet partons and QGP quasiparticles. Then, measuring the jet $p_T$ dependence of the expected bump in the OO/pp ratio of EECs 
would tell us the typical angles of parton deflection due to Moli\`ere scattering in nature. Such measurements 
would constitute direct evidence that energetic jet partons resolve quasiparticle degrees of freedom at sufficiently short distances in an otherwise strongly coupled medium. Such a discovery would offer a window directly into the microscopic structure of QGP: future studies could compare future data to future calculations with varying properties and distributions of QGP quasiparticles, and one can also imagine the possility of learning that the QGP produced in RHIC collisions is more strongly coupled than that produced at the LHC by seeing a distinctive bump in the EEC 
as in Fig.~\ref{fig:eec peaks} in OO collisions at the LHC and seeing it melt away in OO collisions at RHIC.


\textit{Acknowledgments.} --- We thank Cristian Baldenegro, Brian Cole, Giuliano Giacalone, Zachary Hulcher, Gian Michele Innocenti, Yen-Jie Lee, Riccardo Longo, Gunther Roland, Martin Rybar, Anne Sickles, and Adam Takacs for useful discussions. 
Research supported in part by the U.S.~Department of Energy, Office of Science, Office of Nuclear Physics under grant Contract Number DE-SC0011090.
DP acknowledges support from the Ram\'on y
Cajal fellowship RYC2023-044989-I. ASK is supported by the National Science Foundation Graduate Research Fellowship Program under 
Grant No.~2141064. DP and KR are grateful to the Kavli Institute for Theoretical Physics (KITP) for hospitality and support as this work was completed; this research was supported in part by grant NSF PHY-2309135 to the KITP.


\bibliography{bibliography}

\clearpage

\begin{widetext}
\input{supplement}

\clearpage
\end{widetext}


\end{document}

%% file: supplement.tex

\section{Supplemental Material}

\subsection{Determination of $\kappa_{\rm sc}$}

In the Hybrid Model, we assume that jet partons traveling through strongly coupled quark-gluon plasma with a local temperature $T$ lose energy to the plasma at a rate $dE/dx$ that takes the same form as that for an energetic massless parton propagating through the strongly coupled plasma of ${\cal N}=4$
supersymmetric Yang-Mills (SYM) theory with temperature $T$, which can be calculated holographically and is given 
by~\cite{Chesler:2014jva,Chesler:2015nqz}
\begin{equation}
\label{eq:elossrate2}
\frac{dE}{dx} = -\frac{4}{\pi}\frac{E_{\rm in}}{x_{\rm stop}}
\frac{x^2}{x_{\rm stop}^2}\frac{1}{\sqrt{1-(x/x_{\rm stop})^2}}\ , 
\quad {\rm with}\ 
x_{\rm stop} = \frac{E_{\rm in}^{1/3}}{2\,\kappa_{\rm sc}\,T^{4/3}} ,
\end{equation}
where $E_{\rm in}$ is the initial energy of the parton, $x$ is the distance that it has traveled, 
and $x_{\rm stop}$ is the distance over which the initially energetic massless parton would lose all its energy and thermalize unless it splits first. The dimensionless parameter $\kappa_{\rm sc}$ governs the strength of the interaction between 
the massless parton and the strongly coupled medium.
For a massless parton in the fundamental representation of the $SU(N_c)$ gauge group in the strongly coupled plasma of ${\cal N}=4$ SYM theory with large $N_c$ and large  't Hooft coupling $\lambda=g^2 N_c$, one can 
calculate $\kappa_{\rm sc} = 1.05\lambda^{1/6}$~\cite{Chesler:2014jva,Chesler:2015nqz}.
Massless partons in the adjoint representation 
also lose energy as
described by Eq.~\eqref{eq:elossrate2}, but with a larger $\kappa_{\rm gluon} = (C_A/C_F)^{1/3} \kappa_{\rm sc}$~\cite{Gubser:2008as}, meaning that in the Hybrid Model when we describe the energy loss of gluons in parton showers
we choose $\kappa_{\rm gluon}=(9/4)^{1/3}\kappa_{\rm sc}$~\cite{Casalderrey-Solana:2014bpa}. 
Because we are interested in the strongly coupled QGP produced in heavy ion collisions, which is
described by QCD not by ${\cal N}=4$ SYM theory,
in the Hybrid Model we treat $\kappa_{\rm sc}$ as a parameter to be fixed by fitting Hybrid Model calculations of jet observables to experimental data. 
We expect and find~\cite{Casalderrey-Solana:2014bpa,Casalderrey-Solana:2015vaa,Casalderrey-Solana:2016jvj,Casalderrey-Solana:2018wrw}
that the fitted value of $\kappa_{\rm sc}$ for QCD is smaller (in fact by a factor of $\sim 4$) than that for the ${\cal N}=4$ SYM plasma at the same $T$ and $\lambda$ because QCD has fewer degrees of freedom.
As in all previous Hybrid Model studies, we shall only use data from PbPb collisions
when fitting the value of $\kappa_{\rm sc}$.
%
In this Letter, we have made three improvements to the Hybrid Model unrelated to the addition of Moli\`ere scattering, meaning that we need to revisit our choices of 
$\kappa_{\rm sc}$, both for our calculations without and with Moli\`ere scatterings.  
In this Supplemental Material, we explain how we have made these choices.

\begin{figure*}
    \centering
    \begin{subfigure}[t]{0.45\textwidth}
        \centering
        \includegraphics[width=\textwidth]{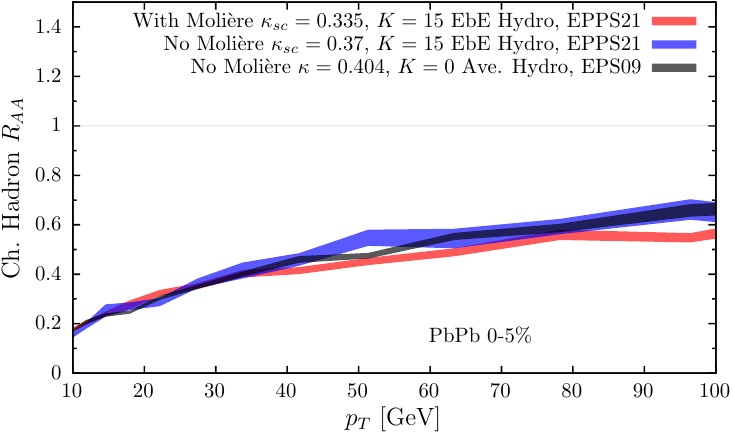}
    \end{subfigure}%
    ~ 
    \begin{subfigure}[t]{0.45\textwidth}
        \centering
        \includegraphics[width=\textwidth]{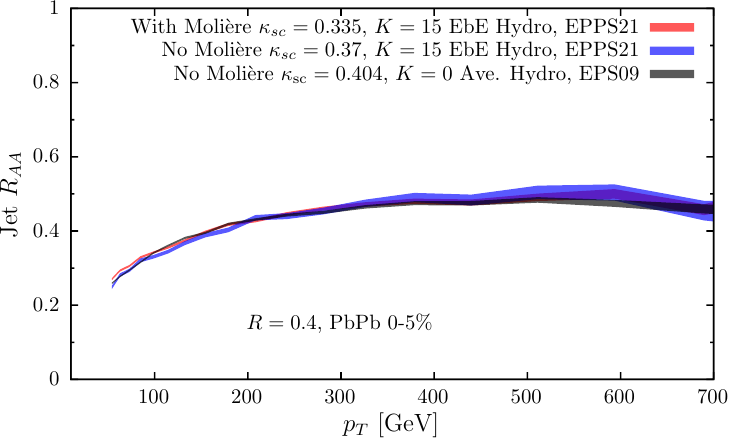}
    \end{subfigure}
    \caption{
    Suppression $R_{\rm AA}$ of charged hadrons (left panel) and jets reconstructed with anti-$k_t$ $R_{\rm jet}=0.4$ (right panel) in 0--5\% central PbPb collisions with collision energy $\sqrt{s_{\rm NN}} = 5.02$ TeV. The black curves show Hybrid Model results using event-averaged hydrodynamic backgrounds~\cite{Shen:2014vra}, with $\kappa_{\rm sc}^{\rm No\,Moli\grave{e}re}=0.404$, as fitted to data in Ref.~\cite{Casalderrey-Solana:2018wrw}, before Moli\`ere scatterings were included in the model and with no soft Gaussian transverse momentum broadening, i.e.~with $K=0$. The blue and red curves show Hybrid Model calculations using event-by-event hydrodynamic backgrounds~\cite{Mantysaari:2025tcg} and including soft Gaussian transverse momentum broadening with $K=15$, without (blue) and with (red) Moli\`ere scatterings. To obtain the blue curves, with no Moli\`ere scattering, we have chosen $\kappa_{\rm sc}^{\rm No\,Moli\grave{e}re}=0.37$. To obtain the red curves, in our calculations that include Moli\`ere scttering we have chosen $\kappa_{\rm sc}^{\rm With\,Moli\grave{e}re}=0.335$.
    As we describe in this Supplemental Material, we have made these choices so as to bring  the blue and red curves into reasonable agreement with the black curves in both panels.  Note also that the old results in the black curves used EPS09 nPDFs~\cite{Eskola:2009uj} while the new results in the blue and red curves use EPPS21 nPDFs~\cite{Eskola:2021nhw}.}
    \label{fig:raa-pbpb}
\end{figure*}

In Ref.~\cite{Casalderrey-Solana:2018wrw},
by fitting Hybrid Model calculations without Moli\`ere scatterings to experimental measurements of
the suppression $R_{\rm AA}$ for charged hadrons and jets in PbPb collisions that were then available,
we obtained the value $\kappa_{\rm sc}=0.404$.
In those calculations and all subsequent Hybrid Model calculations before this work,
jet production was calculated using
EPS09 nuclear PDFs (nPDFs)~\cite{Eskola:2009uj},
and the resulting parton showers described by PYTHIA 8
were then embedded in boost-invariant 
event-averaged hydrodynamic simulations~\cite{Shen:2014vra} of the expanding cooling plasma produced in PbPb collisions. 
These hydrodynamic simulations reproduce the multiplicity of charged particles produced at mid-rapidity in PbPb collisions at the LHC and were obtained by averaging over the fluctuating initial states of many collisions within a given centrality class. In Hybrid Model calculations, the local properties of the plasma, in particular its temperature $T$ which appears explicitly in the energy loss rate~\eqref{eq:elossrate2} and the local fluid velocity (which is important because Eq.~\eqref{eq:elossrate2} is applied after boosting to the local fluid rest frame),
were determined from these hydrodynamic simulations.
The black curves in Fig.~\ref{fig:raa-pbpb}
show the results of Hybrid Model calculations
of $R_{\rm AA}$ for charged hadrons and jets that we have made upon making all of these choices
in the same way as in Ref.~\cite{Casalderrey-Solana:2018wrw}.

Although event-averaged hydrodynamic profiles (as in the black curves in Fig.~\ref{fig:raa-pbpb})
have been used successfully in prior Hybrid Model calculations of jet quenching in PbPb collisions,
event-by-event fluctuations in the shape of the droplet of QGP are much more significant in 
collisions of smaller nuclei, like oxygen.
As our goal in this Letter is to describe jets
in OO collisions,
it is imperative that here we have 
embedded Hybrid Model parton showers in event-by-event hydrodynamic profiles computed within a state-of-the-art framework~\cite{Mantysaari:2025tcg} that describes the formation and subsequent evolution of the QGP in ultrarelativistic OO or PbPb collisions.
Our focus in this Letter is OO collisions, but first we must use calculations for PbPb collisions to redo the choice of $\kappa_{\rm sc}$.
In addition to improving the Hybrid Model treatment of the hydrodynamic droplets of QGP, in our calculations in this Letter
we have calculated jet production using
EPPS21 nPDFs~\cite{Eskola:2021nhw}.
The third improvement that we have made relative to the calculations of Ref.~\cite{Casalderrey-Solana:2018wrw} in which the value $\kappa_{\rm sc}=0.404$ was obtained from a fit to data
is that in this Letter
%
we have modeled the soft exchange of transverse momentum between jet partons and the medium by 
including soft Gaussian transverse momentum
broadening --- such that a jet parton that travels $\delta x$ after a splitting picks up transverse momentum $\langle k_\perp^2\rangle=K T^3 \delta x$ as first described in Ref.~\cite{Casalderrey-Solana:2016jvj}, where we choose the parameter $K=15$ as 
described in Ref.~\cite{Hulcher:2026dht}. 
These three improvements to the Hybrid Model
necessitate revisiting the choice of $\kappa_{\rm sc}$.

Fig.~\ref{fig:raa-pbpb} shows Hybrid Model calculations of $R_{\rm AA}$ for charged hadrons in the left panel and for jets with anti-$k_t$ radius $R_{\rm jet}=0.4$ in the right panel in $0$--$5\%$ central PbPb collisions with a collision energy of $\sqrt{s_{\rm NN}}=5.02$ TeV. As noted already, the black curves
correspond to calculations made with the same choices as in 
Ref.~\cite{Casalderrey-Solana:2018wrw} --- event-averaged hydrodynamic backgrounds, EPS09 nPDFs,  $K=0$, and no Moli\`ere scatterings --- and with $\kappa_{\rm sc}=0.404$, as determined in 
Ref.~\cite{Casalderrey-Solana:2018wrw}.
The blue and red curves show Hybrid Model calculations performed by embedding parton showers generated using PYTHIA 8 with EPPS21 nPDFs~\cite{Eskola:2021nhw} into event-by-event hydrodynamic backgrounds~\cite{Mantysaari:2025tcg}, with soft Gaussian transverse momentum broadening with $K=15$, without (blue curves) and with (red curves) Moli\`ere scatterings. 
In future work, the value of $\kappa_{\rm sc}$ should be chosen via a Bayesian analysis of many data sets that simultaneously constrain $\kappa_{\rm sc}$, $K$ 
and the two parameters $g_s$ and $a$ that govern Moli\`ere scattering.
In this Letter, we are not aiming to determine the values of these parameters with reliable quantification of uncertainties. Our goal is the initial exploration of jet substructure observables in OO collisions for the purpose of identifying observables that are sensitive to Moli\`ere scattering and highlighting possible paths toward distinctive and model-independent detections of hard scattering of jet partons off QGP quasiparticles. If and when experimental data yields such detections, this will provide central inputs to a future Bayesian quantification of the uncertainty in the Hybrid Model parameters.
For our present purposes, a simpler approach suffices. For this study, we have chosen $K=15$, $g_s=2.25$ and $a=10$ as described in our Letter.
Here, we use the Hybrid Model calculations
of charged hadron and jet suppression in PbPb
collisions in Fig.~\ref{fig:raa-pbpb} to motivate
choosing
$\kappa_{\rm sc}^{\rm No\,Moli\grave{e}re}=0.37$ 
in the absence of Moli\`ere scatterings, as in the blue curves in Fig.~\ref{fig:raa-pbpb}, 
and $\kappa_{\rm sc}^{\rm With\,Moli\grave{e}re}=0.335$ in our calculations that include Moli\`ere scatterings, as in the red curves.
We have chosen these values of $\kappa_{\rm sc}$
so that the red curves and blue curves in Fig.~\ref{fig:raa-pbpb} are similar to the black curves.

We note that the red curve in the left panel of Fig.~\ref{fig:raa-pbpb} is slightly below the black curve for $p_T^{\rm ch\text{-}had} \gtrsim 40$ GeV. 
We have prioritized agreement 
for $p_T^{\rm ch\text{-}had}\lesssim 40$~GeV as that is where the error bars in the experimental data employed in the fit of Ref.~\cite{Casalderrey-Solana:2019ubu} are smallest.
Also, if we were to
increase the value of $\kappa_{\rm sc}^{\rm With\,Moli\grave{e}re}$ 
so as to improve the agreement between the red and black curves at higher $p_T$, this would 
worsen the agreement between the red and black curves in the right panel 
of Fig.~\ref{fig:raa-pbpb}, meaning that it would
worsen the fit to jet $R_{\rm AA}$.
Since in our Letter we wish to study the effects of including/excluding Moli\`ere scatterings on the internal structure of jets, it is particularly important that we compare samples of jets with/without Moli\`ere scattering that have experienced the same amount of suppression, as quantified by the jet $R_{\rm AA}$ observable. Hence, we prioritize the fit to jet $R_{\rm AA}$ (right panel) over the fit to charged-hadron $R_{\rm AA}$ (left panel). 
In the future, and in particular once experimental measurements of the OO jet substructure observables whose importance we have highlighted in this Letter are in hand, the values of $\kappa_{\rm sc}$ and the other Hybrid Model parameters should be constrained via a Bayesian analysis of a large suite of experimental data.


We stress that we have chosen values of $\kappa_{\rm sc}$ for our Hybrid Model calculations with and without Molière scattering solely via calculations of charged hadron and jet suppression in central PbPb collisions.  We have not employed any jet substructure observables, and we have not employed any calculations of jet quenching observables in OO collisions.



\subsection{Jet Suppression in OO Collisions}

\begin{figure*}
    \centering
    \begin{subfigure}[t]{0.45\textwidth}
        \centering
        \includegraphics[width=\textwidth]{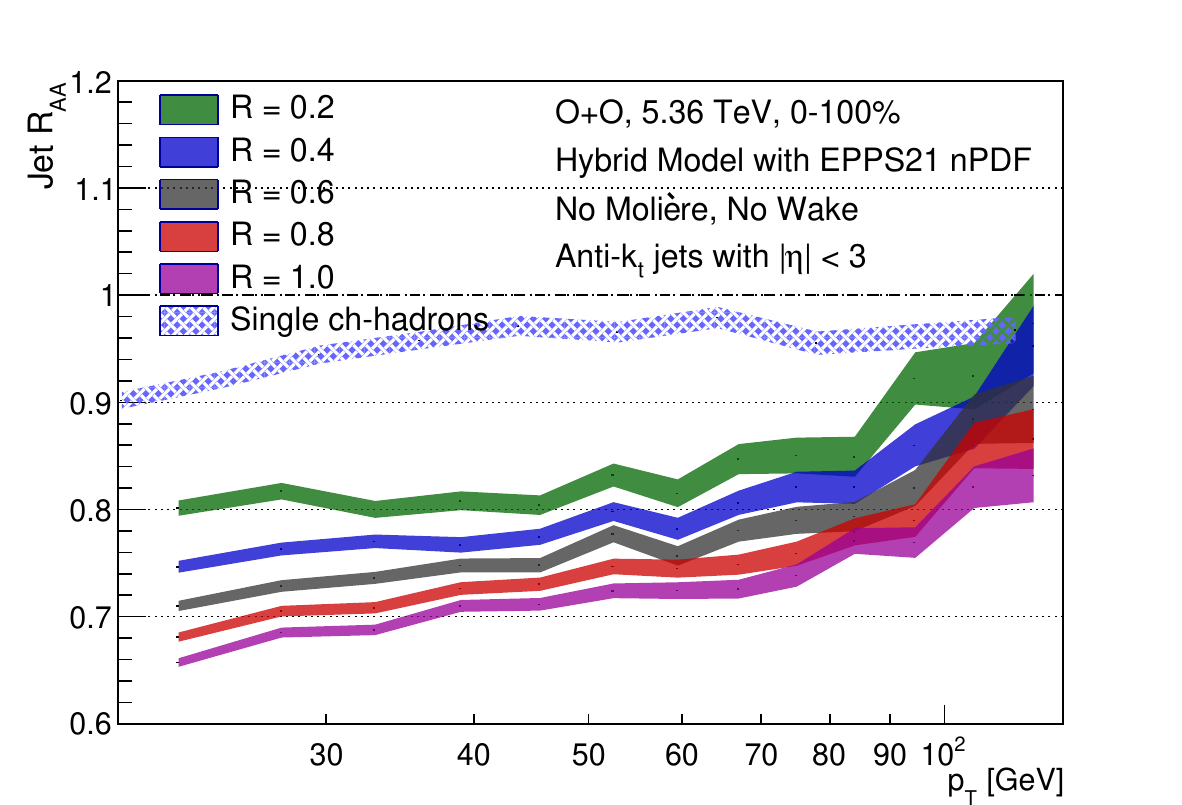}
    \end{subfigure}%
    ~ 
    \begin{subfigure}[t]{0.45\textwidth}
        \centering
        \includegraphics[width=\textwidth]{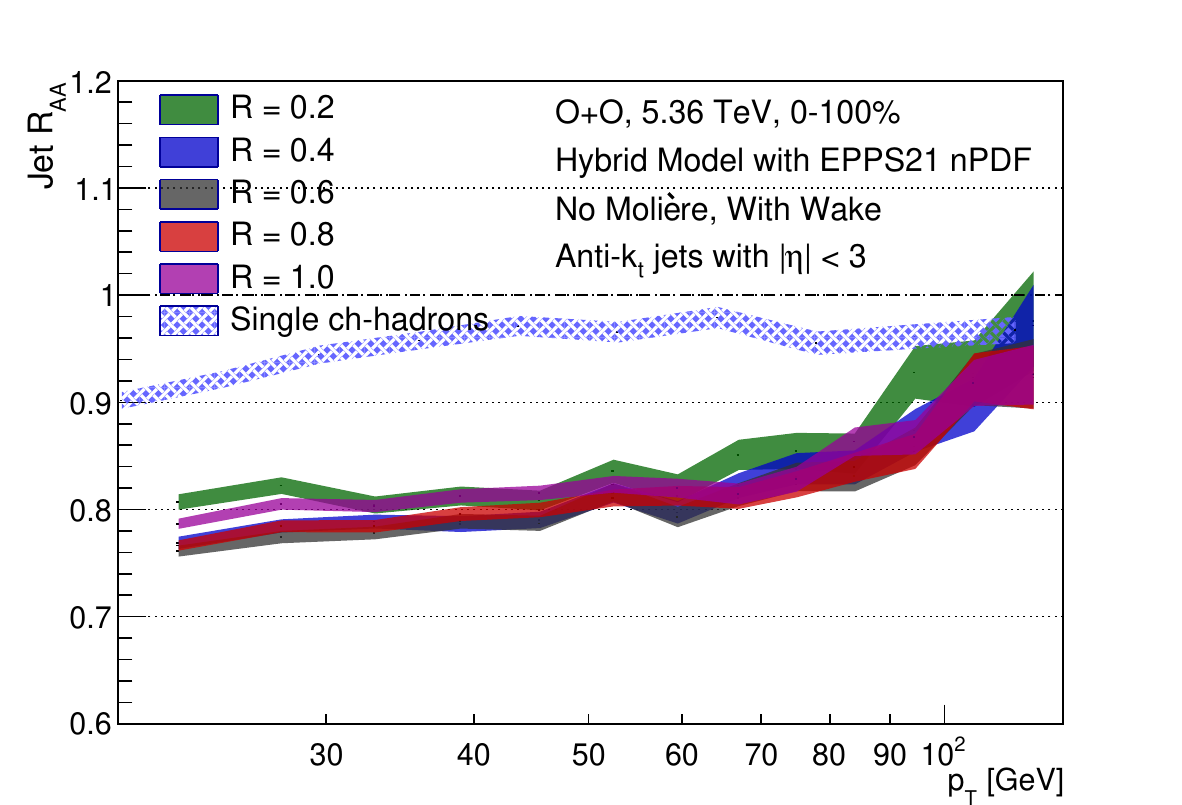}
    \end{subfigure}%

    \begin{subfigure}[t]{0.45\textwidth}
        \centering
        \includegraphics[width=\textwidth]{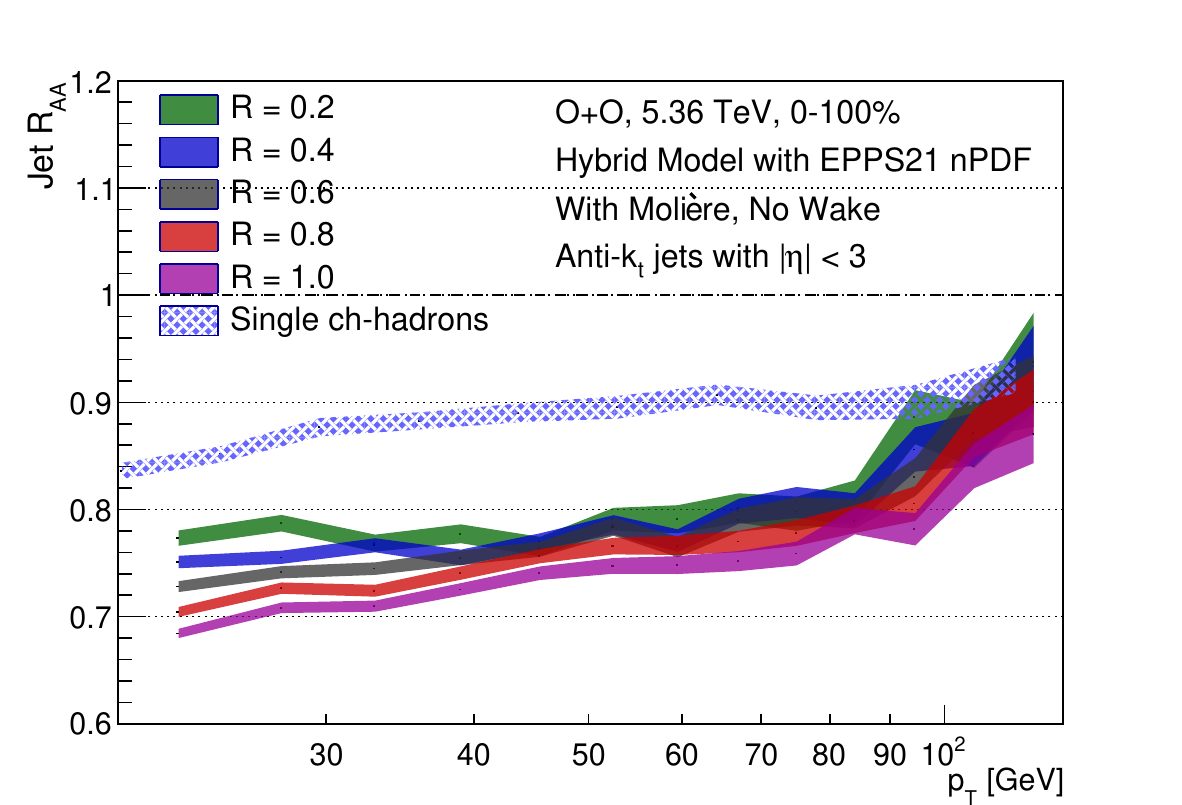}
    \end{subfigure}%
    ~ 
    \begin{subfigure}[t]{0.45\textwidth}
        \centering
        \includegraphics[width=\textwidth]{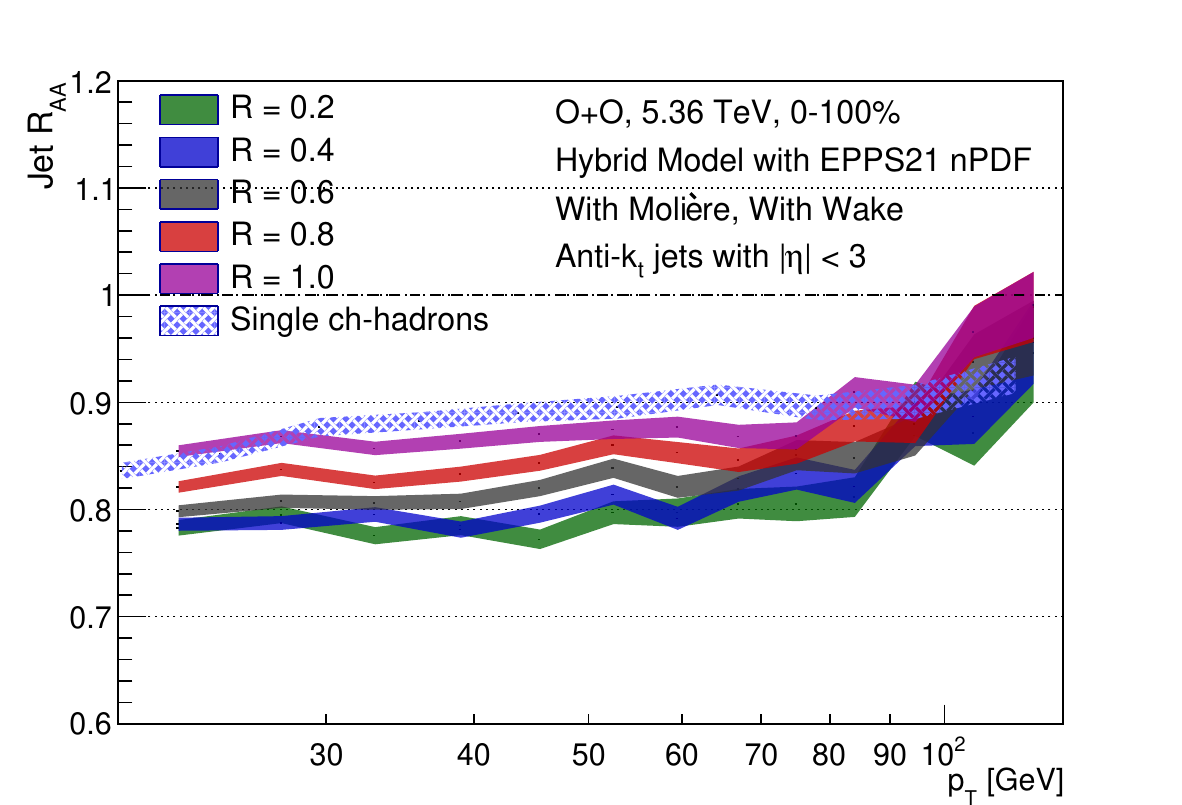}
    \end{subfigure}
    \caption{Solid colored bands: Hybrid Model calculations of $R_{\rm AA}$ for anti-$k_t$ jets with radii $R_{\rm jet} = 0.2$, 0.4, 0.6, 0.8, and 1.0, with $|\eta| < 3$, as a function of $p_T$ in minimum-bias OO collisions with beam energy $\sqrt{s_{\rm NN}}=5.36$~TeV. Hatched blue band: Hybrid Model calculations of $R_{\rm AA}$ for charged hadrons with $|\eta| < 3$ as a function of $p_T$ in the same sample of OO collisions. The effects of Moli\`ere scattering are excluded from the upper panels, and are included in the lower panels. The effects of jet wakes are excluded from the left panels and are included in the right panels. The hatched blue band in the bottom-right panel is similar (differing only in the $|\eta|$ cut) to the red band in Fig.~\ref{fig:ch-had-raa} that we have compared to CMS data. Future experimental measurements of jet $R_{\rm AA}$ for jets with varying $R_{\rm jet}$ can be compared to the solid colored bands in the bottom-right panel. The other three panels show how jet $R_{\rm AA}$ changes when we turn off either jet wakes or Moli\`ere scattering or both, which cannot be done in experimental data.
        }
    \label{fig:jet-raa}
\end{figure*}

In Fig.~\ref{fig:ch-had-raa} of this Letter, 
we have compared Hybrid Model calculations 
of $R_{\rm AA}$ for charged hadrons in OO collisions, and compared them to recent measurements from CMS~\cite{CMS:2025bta}.  
Since a high-$p_T$ charged hadron is likely the leading hadron from a jet, this is a standard 
observable via which to quantify jet suppression.
We have seen in Fig.~\ref{fig:ch-had-raa} that in the Hybrid Model nPDFs, strongly coupled energy
loss, and Moli\`ere scattering all have comparable effects on this observable and have seen 
that the Hybrid Model
describes this measure of jet suppression in OO collisions well only when we include Moli\`ere scatterings.
The ATLAS collaboration has also recently
reported measurements of the dijet asymmetry
in central and peripheral OO collisions~\cite{ATLAS:2025ooe}.
This is also a standard observable via which to quantify jet suppression, and has the advantage that it is less sensitive to nPDF effects 
than $R_{\rm AA}$.  We have checked, however,
that in the Hybrid Model this observable also has little sensitivity to whether we do or do not include Moli\`ere scattering, which makes it of less interest for our purposes in this Letter.


In this Supplemental Material we report Hybrid Model calculations of
a third standard observable via which to quantify jet suppression, namely $R_{\rm AA}$ for jets
themselves, in OO collisions.
Fig.~\ref{fig:jet-raa} shows the nuclear modification factor $R_{\rm AA}$ for anti-$k_t$ jets in OO collisions, for jet radii $R_{\rm jet} = 0.2, \, 0.4, \, 0.6, \, 0.8,$ and $1.0$, with $|\eta| < 3$, as a function of jet $p_T$. In the hashed bands, we also show calculations of $R_{\rm AA}$ for single high-$p_T$ charged hadrons with $|\eta| < 3$. The results in the top (bottom) panels exclude (include) the effects of Moli\`ere scattering of jet partons off quasiparticles in the medium. The results in the left (right) panels exclude (include) the hadrons formed via the freezeout of jet wakes. 
Although only the calculations in the bottom-right panel of Fig.~\ref{fig:jet-raa} should be compared to future
experimental measurements, since in nature unlike in the Hybrid Model it is impossible to turn physical effects like jet wakes or hard scattering off, comparison among the four panels is very instructive.



We can start in the top-left panel, with no Moli\`ere scattering and no hadrons originating from jet wakes.  
Here, and in fact in all the panels, we
see that the jet $R_{\rm AA}$ is more suppressed
than $R_{\rm AA}$ for charged hadrons across
most of the $p_T$ range shown and for jet radii $R_{\rm jet} \lesssim 1$. 
This is because a jet with a given $p_T$ 
originates from a shower containing multiple partons,
meaning more resolved sources of energy loss,
than is the case for just one hadron --- which is typically the leading hadron in a jet, originating from a single leading parton in the shower.
The ordering of the suppression of jet $R_{\rm AA}$
with jet radius $R_{\rm jet}$ seen most clearly 
in the top-left panel in the absence of jet wakes and Moli\`ere scattering can be understood similarly.
%
%
Jets with larger radii are systematically more suppressed than jets with smaller radii. This is because wider jets with a given $p_T$ contain more resolved sources of energy loss than skinnier jets, and therefore will be more suppressed than the skinnier 
jets~\cite{Milhano:2015mng,
Rajagopal:2016uip,Casalderrey-Solana:2016jvj,Brewer:2017fqy,Hulcher:2017cpt,
Mehtar-Tani:2017web,Casalderrey-Solana:2018wrw,Casalderrey-Solana:2019ubu,Caucal:2019uvr,Pablos:2019ngg,
Du:2020pmp,
Caucal:2021cfb,
Brewer:2021hmh,Pablos:2022mrx,Hulcher:2022kmn,ATLAS:2023hso, Kudinoor:2025ilx, ATLAS:2025lfb, Kudinoor:2025gao,Tachibana:2025rcx}. 
This effect is present in all panels, but is most apparent in the top-left panel because other physical phenomena have significant effects in other panels.

When jet wakes are included in our calculations the energy and momentum which is lost by each jet parton and which is deposited into the plasma in the form of wakes is partially recovered during jet reconstruction.
Comparing the top-right panel of Fig.~\ref{fig:jet-raa} to the top-left panel, we see that including hadrons originating from jet wakes
has little effect on the suppression of $R_{\rm jet}=0.2$ jets --- reconstructing such skinny jets does not
catch much wake ---
but with increasing jet radius more and more of the hadrons from jet wakes 
are found within the jet radius, pushing $R_{\rm AA}$ upwards by more and more for larger and larger $R_{\rm jet}$. 
As a result, in the top right panel of Fig.~\ref{fig:jet-raa}, we see that $R_{\rm AA}$ 
hardly depends on the jet radius when jet wakes are included and Moli\`ere scattering is excluded. 
The fact that all the colored bands in the top-right panel are so close to each other arises from a near cancellation of the
increase in suppression with increasing $R_{\rm jet}$ because the parton showers in wider jets lose more energy and the decrease in suppression with increasing $R_{\rm jet}$ because wider jets include more of the hadrons from the freezeout of jet wakes~\cite{Pablos:2019ngg,Mehtar-Tani:2021fud}.


Comparing the bottom-left panel to the top-left panel, we see that turning on Moli\`ere scattering increases the suppression of the skinniest jets because it kicks partons out of the jet cone while it reduces the suppression of the wider jets because in this case the partons from Moli\`ere scattering, which constitute additional sources of energy loss, are more likely to remain in the jet cone.  The effect is to squeeze the colored bands in the bottom-left panel closer together.
When we then include the hadrons from jet wakes in the bottom-right panel, pushing the $R_{\rm AA}$ for wider and wider jets more and more upward as we have already discussed, the effect is to invert the order of the colored bands.  Only in the bottom-right panel do we see less suppression for wider
jets, so much so that in this panel, with all physical effects included, 
the $R_{\rm AA}$ of $R_{\rm jet} = 1.0$ jets is similar to $R_{\rm AA}$ of charged hadrons. This is a coincidence due to a near-cancellation among three effects: parton energy loss, jet wakes, and Moli\`ere scattering. 
Namely, the suppression of $R_{\rm jet} = 1.0$ jets at a given $p_T$ due to parton energy loss happens to (almost) be canceled by the enhancement of $R = 1.0$ jets at that same jet $p_T$ due to the effects of jet wakes and Moli\`ere scattering. 
A (near) cancellation like this between contributions from differing physical phenomena is certainly a model-dependent result, as it depends on the quantitative correctness of the calculation of each. We note in this regard that
the Hybrid Model treatment of the soft hadrons originating from jet wakes is crude, with weaknesses that were enumerated already in Ref.~\cite{Casalderrey-Solana:2016jvj} that are particularly significant at the largest angles away from the jet axis,
and that 
have motivated work towards an improved treatment begun in Ref.~\cite{Casalderrey-Solana:2020rsj}. 
Once the improved treatment of the wake 
that is being developed by the authors of Ref.~\cite{Casalderrey-Solana:2020rsj} is implemented in the Hybrid Model, the effects of jet wakes in the right panels of Fig.~\ref{fig:jet-raa} will need to be reassessed, as will the near cancellation that we have discussed.


It is apparent from Fig.~\ref{fig:jet-raa} that the suppression $R_{\rm AA}$ of jets with varying radii $R_{\rm jet}$ is
sensitive to the effects of Moli\`ere scattering.
However, it is also apparent that it is comparably sensitive to parton energy loss and jet wakes, and that these different phenomena yield differing and competing $R_{\rm jet}$-dependence of the suppression. 
This makes it challenging to 
attempt to draw distinctive conclusions about any one of these phenomena from experimental measurements of this observable alone.

\subsection{EECs in OO Collisions Calculated for Different Collision Centralities}

\begin{figure*}
    \centering
    \begin{subfigure}[t]{0.45\textwidth}
        \centering
        \includegraphics[width=\textwidth]{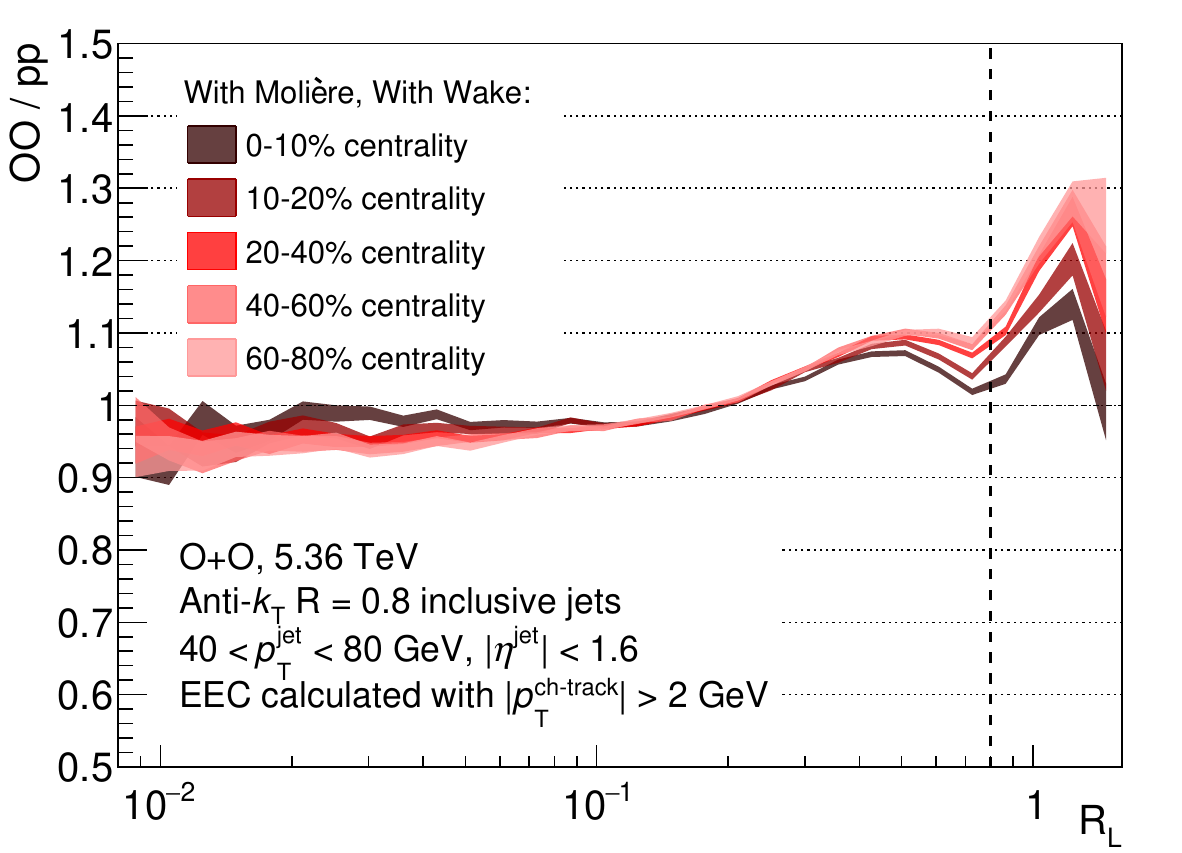}
    \end{subfigure}%
    ~ 
    \begin{subfigure}[t]{0.45\textwidth}
        \centering
        \includegraphics[width=\textwidth]{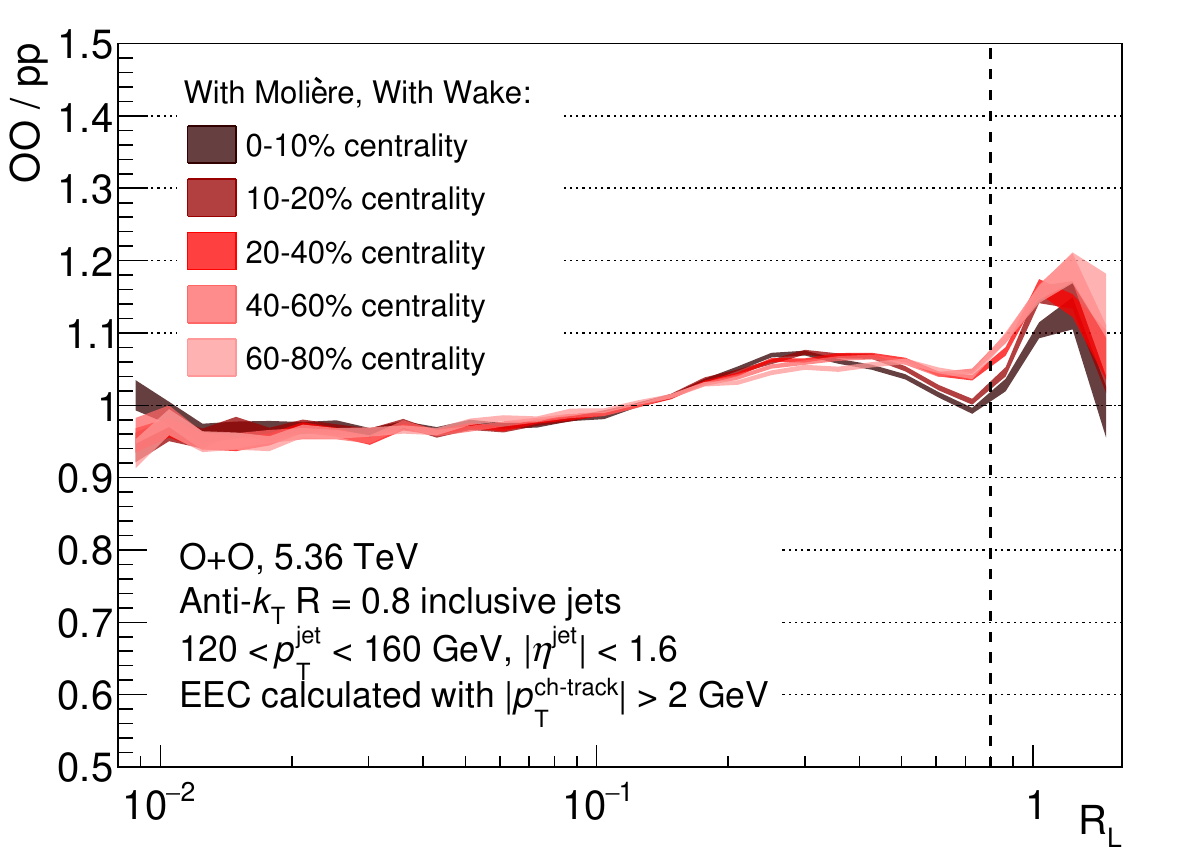}
    \end{subfigure}
    
    \caption{
    Hybrid Model calculations of OO/pp ratios of EECs calculated for $R_{\rm jet} = 0.8$ jets with $40 < p_T < 80$ GeV (left) and $120 < p_T < 160$ GeV (right), for different centrality classes, as a function of $R_L$. All calculations include Moli\`ere scattering and hadrons originating from jet wakes. EECs are calculated using charged-particles with $p_T>2\, {\rm GeV}$, which means that hadrons from jet wakes make a negligible contribution to this observable. The dashed vertical line in each panel corresponds to $R_L = R_{\rm jet}$.  
    } 
    \label{fig:oo eec bump vary centrality}
\end{figure*}

In Figs.~\ref{fig:eec min bias 40-80} and \ref{fig:eec peaks} of this Letter, we have studied how Moli\`ere scatterings and jet wakes modify the distributions of EEC$(R_L)$ in minimum-bias ($0-100\%$ centrality) OO collisions. In this Supplemental Material, we present Hybrid Model calculations of EEC$(R_L)$ for OO collisions with differential ranges of collision centralities.
Fig.~\ref{fig:oo eec bump vary centrality} shows the OO/pp ratios of EEC$(R_L)$ for anti-$k_t$ $R_{\rm jet}=0.8$ jets with $40 < p_T^{\rm jet} < 80$ GeV (left panel) and $120 < p_T^{\rm jet} < 160$ GeV (right panel). The EECs are calculated using charged-particle tracks with $p_T^{\rm ch~track} > 2$ GeV with jet wakes and Moli\`ere scattering both included. (However, the restriction $p_T^{\rm ch~track} > 2$ GeV means that contributions from jet wakes are almost completely negligible.) From darkest to lightest shade, the colored bands in each panel correspond to OO collisions in the centrality ranges $0-10\%$, $10-20\%$, $20-40\%$, $40-60\%$, and $60-80\%$.  

The left (right) panel of Fig.~\ref{fig:oo eec bump vary centrality}
should be compared to the results of Hybrid Model calculations for minimum-bias (0-100\% centrality) OO collisions shown in the pink (reddish-brown) 
bands in Fig.~\ref{fig:eec peaks}. In our discussion of Fig.~\ref{fig:eec peaks}, we have highlighted the broad ``bump'' in the EEC ratio
that is at around $R_L\sim 0.5$ for jets with 
$40<p_T^{\rm jet}<80$~GeV and around $R_L\sim 0.3$ for jets with $120<p_T^{\rm jet}<160$~GeV. The calculations in Fig.~\ref{fig:eec peaks} were done for minimum-bias (0-100\% centrality) OO collisions; we see by comparing these results to those in the darkest bands in Fig.~\ref{fig:oo eec bump vary centrality} that the prominence of this bump only increases slightly if we consider 
0-10\% central OO collisions.  (The height of the bump does not change appreciably, but the depth of the dip to the right of the bump increases slightly.)  This indicates that it is not necessary to select a sample of collisions with 0-10\% centrality in order to measure this distinctive signature of Moli\`ere scattering in the OO/pp EEC ratio. Fig.~\ref{fig:oo eec bump vary centrality} also confirms that selecting a sample of collisions with some more peripheral differential range of collision centralities also presents no significant advantage. 

We conclude this Supplemental Material with a few words about the apparent peak in the OO/pp EEC ratio for $R_{\rm jet}<R_L<2R_{\rm jet}$ whose position in $R_L$ is governed by the value of $R_{\rm jet}$, not by any angular scale associated with Moli\`ere scattering. As $R_L$ increases beyond $R_{\rm jet}$, the two point EEC is a correlation between the energy flow in two directions neither of which can be in the central core of the jet.  Such correlations are small in pp collisions and decrease rapidly with increasing $R_L$. The OO/pp EEC ratio
is enhanced in OO collisions because the denominator is dropping and
because Moli\`ere scattering and jet wakes populate the halos of jets with some energy. With the track cut
$p_T^{\rm ch~track}>2$~GeV that we are employing in our analysis, though, only Moli\`ere scattering is relevant.  In this sense, the rise in the OO/pp EEC ratio for $R_L>R_{\rm jet}$, calculated with the track cut, 
is a direct consequence of Moli\`ere scattering.  However, what cuts this rise off and turns OO/pp EEC ratio back down towards unity is a purely kinematic effect that has nothing to do with Moli\`ere scattering: as $R_L$ approaches $2 R_{\rm jet}$, there are fewer and fewer pairs of charged-particle tracks within the jet cone that are separated by $R_L$. For further discussion of  edge effects in EECs, see Ref.~\cite{Andres:2025yls}.